\documentclass[11pt,english]{article}
 \pdfoutput=1
\usepackage[maccyr]{inputenc}
\usepackage{amssymb,amsmath}
\usepackage{cite}
\usepackage{color}
\usepackage{bbm}
\usepackage{setspace}
\usepackage{empheq}
\usepackage{enumerate}
\usepackage{amsthm}
\def\hybrid{
        \topmargin -20pt
        \oddsidemargin 0pt
        \headheight 0pt \headsep 0pt
        \textwidth 6.25in 
        \textheight 9.5in 
        \marginparwidth .875in
        \parskip 5pt plus 1pt \jot = 1.5ex}

\hybrid

 \def\det{{\rm det\,}}
\def\be{\begin{equation}}
\def\ee{\end{equation}}
\def\bea{\begin{eqnarray}}
\def\eea{\end{eqnarray}}
\def\ba{\begin{align}}
\def\ea{\end{align}}

\begin{document}
\begin{titlepage}
\samepage{
\setcounter{page}{1}
\vspace{-1cm}
\begin{flushright}
{\small
CPHT-RR022.052017
\\
ROM2F/2017/03 \\
}
\end{flushright}
\begin{center}
\vspace{0.4cm}
{\Large \bf  Open Strings and Electric Fields in Compact Spaces}

\vspace{15mm}
\begin{minipage}{0.97\linewidth}
\begin{center}
{\normalsize\bf Cezar Condeescu$^{1}$,~ Emilian Dudas$^{2}$ and Gianfranco Pradisi$^{3,4}$\\
}
\vspace{10mm}

\end{center}
\vspace{.2cm} \hspace{2.0cm}
\begin{minipage}{1.\linewidth}
\begin{small}
\begin{itemize}
	\item[$^{1}$]   Department of Theoretical Physics \\
 ``Horia Hulubei" National Institute of Physics and Nuclear Engineering \\
 P.O. Box MG-6, M\u{a}gurele - Bucharest, 077125, Jud. Ilfov, Rom\^{a}nia    
	 \item[$^{2}$]  Centre de Physique Th\'eorique, \'Ecole Polytechnique, CNRS \\ Universit\'e Paris-Saclay, F-91128 Palaiseau, France   
	\item[${}^3$]  Dipartimento di Fisica, Universit\`a di Roma ``Tor Vergata"\\
          Via della Ricerca Scientifica 1, 00133 Roma, Italy
          \item[${}^4$]  INFN, Sezione di Roma ``Tor Vergata"\\
          Via della Ricerca Scientifica 1, 00133 Roma, Italy
\end{itemize}
\end{small}
 \vspace{0.2cm}
 \end{minipage}
\end{minipage}

\vspace{10mm}
\begin{abstract}\vspace{1mm}
{\normalsize We analyse open strings with background electric fields in the internal space, T-dual to branes moving with constant velocities in the internal space. We find that the direction of the electric fields inside a two torus, dual to the D-brane velocities, has to be quantised such that the corresponding direction is compact. This implies that D-brane motion in the internal torus is periodic, with a periodicity that can be parametrically large in terms of the internal radii.  By S-duality, this is mapped into an internal magnetic field in a three torus,  a quantum mechanical analysis of which yields a similar result, {\it i.e.} the parallel direction to the magnetic field has to be compact. Furthermore, for the magnetic case, we find the Landau level degeneracy as being given by the greatest common divisor of the flux numbers. We carry on the string quantisation and derive the relevant partition functions for these models. Our analysis includes also the case of oblique electric fields which can arise when several stacks of branes are present. Compact dimensions and/or oblique sectors influence the energy loss of the system through pair-creation and thus can be relevant for inflationary scenarios with branes. Finally, we show that the compact energy loss is always larger than the non-compact one. 
}
\end{abstract}
\end{center}

\smallskip}

\vfill
\begin{center}
Emails: ccezar@theory.nipne.ro, emilian.dudas@cpht.polytechnique.fr,\\ \ \ \ gianfranco.pradisi@roma2.infin.it.
\end{center}

\end{titlepage}

\tableofcontents

\section{Introduction}

Open strings \cite{orientifolds} can be quantised exactly in a constant electromagnetic field background \cite{Fradkin:1985qd,Abouelsaood:1986gd,Nesterenko:1989pz}. The case of magnetic fields and their T-dual version of branes at angles have been widely studied in the literature starting from \cite{Bachas:1995ik} due to their promising phenomenological features of realising Standard Model like gauge groups on magnetised/intersecting D-branes, while also preserving $N=1$ supersymmetry \cite{intersecting,orientifolds}. On the other hand, open string models with electric fields, that were pioneered in \cite{Bachas:1992bh}, have received far less attention due to the fact that supersymmetry is always broken in the charged sectors, resulting in systems that are in principle unstable. However, they can offer exact CFT models for studying D-brane dynamics, as for example in \cite{bachas_dynamics, ambjorn,bachashull,cristina,pioline, mcallister}. An important application would be to inflationary cosmology \cite{inflation} where, in the T-dual version of moving branes, one or more positions of branes are identified with the inflaton(s). 

Our work focuses on open strings with background constant electric fields in toroidal compactifications.  It is well known that magnetic fields in compact spaces have to satisfy Dirac quantisation conditions. This is no longer true for electric fields at a perturbative level, due to the fact that one of the legs of the field strength lies always in the non-compact time direction. As we will show, there are non-perturbative quantisation conditions for the components of the electric field along the torus axes, arising from the gauge invariance of $U(1)$ Wilson loops that force the corresponding components of the gauge potential to be compact variables. These conditions have a simple interpretation in the T-dual version as quantisation of momenta of D0 particles along the compact directions. Moreover, from the non-perturbative consistency  one can extract a quantisation condition for the orientation of the electric field inside the torus that is independent on the string coupling constant and hence could in principle  arise at a perturbative level. We consider the simplest possible case, that of an electric field pointing into a generic direction inside a rectangular two torus. {\it The main results of the paper is that  the direction of the electric field has to be compact}. We show this in various ways. Aside from the non-perturbative argument mentioned above, one can derive the same result by making use of the S-duality between electric and magnetic fields. Furthermore we consider, at the quantum mechanical level, a magnetic field pointing into a generic direction inside a two torus (contained in a three torus) such that the electric field case will be an analytic continuation of the magnetic one. We derive here the degeneracy of the Landau levels, relevant for model building, which turns out to be given by the { \it greatest common divisor of the two non-zero flux numbers}. Dirac quantisation conditions immediately imply that {\it the direction parallel to magnetic field is periodic} and since the allowed string momentum is always parallel one has a quantised momentum as well. However, in one particular gauge we are able to construct wave functions respecting the periodicities of the three torus only in the case when also the coordinate orthogonal to the magnetic field is compact. In turn this further implies that the squared modulus of the complex structure of the torus is fixed to be a rational number. It would be interesting to determine whether this condition is indeed also necessary as it would have important implications for moduli stabilisation. The same analysis for the case of the electric field implies in one gauge that quantum mechanically there is no visible quantisation condition, whereas in a different gauge the direction parallel to the electric field comes out to be compact. In principle, the allowed string momentum modes (always orthogonal to the electric field) may or may not belong to a lattice, depending on whether the direction orthogonal to electric field is compact. 

We also present the quantisation of string models with electric fields in internal spaces and construct explicitly their annulus amplitudes taking into account the quantisation conditions for the orientation of the electric field. Strictly speaking, in order to build a consistent string vacuum, one should also consider the contributions of the M\"obius Strip amplitude for open strings. This  can be done in the usual way by applying the orientifold projection \cite{orientifolds}. Our work extends previous results also in another direction, that of oblique electric fields (for the case of oblique magnetic fields see \cite{Gorlich:2004zs, Antoniadis:2004pp,Bianchi:2005yz, Bianchi:2005sa}). We point out that in models with several stacks of branes the possibility arises of having open strings stretched between (necessarily) different branes with electric fields at an angle. In such a situation, the field strengths at the two boundaries of the string do not commute, leading to a more complicated algebra of zero modes and to a non-linear dependence of the induced electric shift in terms of the `rapidities'. Models with oblique electric fields realise a Thomas precession effect for open strings. In the limit of small electric fields (small velocities in the T-dual version) they reproduce the results of the parallel case and thus are expected to be relevant (only) in the ultra-relativistic limit. 

 Finally, we analyse also the energy loss of D-branes in constant electric fields by pair creation. There are two cases that one can compare, depending on the compactness of the direction orthogonal to the electric field. We show that the compact energy loss is always larger than the non-compact one for any finite values of the radius and electric field shift, the two becoming equal asymptotically. Increasing the radius, the compact energy loss decreases implying also that larger radii would yield a greater number of e-folds in inflationary scenarios with moving branes.

The paper is organised as follows. In Section \ref{s2} and \ref{s3} we discuss the quantisation conditions for electric fields in internal spaces from a non-perturbative and S-dual point of view. Section \ref{s4} contains the quantum mechanical analysis of charged particle in electric and magnetic fields at a generic angle with respect to the torus axes. The boundary conditions for open strings in constant electric fields and the various possible sectors (charged/neutral and parallel/oblique) are considered in    
Section \ref{s5}. Furthermore, the quantisation of these models and the corresponding annulus amplitudes can be found in Section \ref{s6} for the case of parallel electric fields\footnote{We call parallel the situation where the two electric fields at the boundaries of the open string are parallel with one another (or one of them is zero). They can still make an (rational) angle with respect to the torus axes!} and in Section \ref{s7} for the case of oblique electric fields. Finally, we discuss the energy loss by pair creation in Section \ref{s8} and our conclusions are contained in Section \ref{s9}.

\section{Brane Motions and Electric Fields in Internal Spaces}
\label{s2}
Let us consider a D2 brane in an internal torus $(x_4,x_5)$, taking for simplicity to be a square, of radii $R_4,R_5$ respectively, and add an electric field making an angle $\beta$ with $x_4$, i.e.  $F_{04} \equiv E_4= E \cos \beta$, $F_{05} \equiv E_5= E \sin \beta$.
After T-dualities in $x_4,x_5$, one gets a point-like D0 brane moving with a constant velocity ${\bf v} = {\bf E} $ with ${\bf v}  = (v \cos \beta, v \sin \beta)$.
The momenta of the D0 particle along the two internal directions have to be quantised\footnote{The following argument is similar to the one in \cite{bachashull}.}. The corresponding conditions are
\bea
&& p_4 = \frac{T_0 v \cos \beta}{\sqrt{1-v^2}} = \frac{q}{R'_4} \ , \nonumber \\
&& p_5 = \frac{T_0 v \sin \beta}{\sqrt{1-v^2}} = \frac{p}{R'_5}  \ , \label{bm1} 
\eea
where $R'_4,R'_5$ are the two T-dual radii, $T_0$ is the D0 brane tension and $p,q$ are integers. Notice that since $T_0 \sim 1/g_s$, the quantisation conditions are non-perturbative in nature. The direction of the electric field, on the other hand, is also quantised and determined by
\be
\tan \beta =  \frac{p}{q} \frac{R'_4}{R'_5} =  \frac{p}{q} \frac{R_5}{R_4}   \ . \label{bm2}
\ee  
Notice that the string coupling does not appear in the quantisation condition for the direction of the electric field.\\
\ 
The quantisation condition can also be found in a neat way by starting from the Dirac-Born-Infeld (DBI) action 
\be
{\cal L}_{D2} = - T_2 \sqrt{- det (g_{mn} + F_{mn} + B_{mn}) } = - T_2  (2 \pi)^2 R_4 R_5 \sqrt{1 - E_4^2 - E_5^2}  \ . \label{bm3}
\ee
The zero mode (Wilson line) of $A_4$ is a compact variable 
\be 
T_F \oint A_4 dx^4 = 2 \pi R_4 T_F \delta A_4 = 2 \pi \ , \label{bm4}
\ee
where $T_F$ is the fundamental string tension. Therefore $A_4  \sim A_4 + \frac{1}{T_F R_4}$. A similar periodicity
is found for $A_5$. Consequently, the variables conjugated to $A_4,A_5$ 
\bea
&& \Pi_4 = \frac{\delta \cal L}{\delta \dot A_4} = \frac{T_2  (2 \pi)^2 R_4 R_5 E_4}{\sqrt{1 - E_4^2 - E_5^2}} =  T_F q R_4 \ , \nonumber \\
&& \Pi_5 = \frac{\delta \cal L}{\delta \dot A_5} = \frac{T_2  (2 \pi)^2 R_4 R_5 E_5}{\sqrt{1 - E_4^2 - E_5^2}} =  T_F p R_5  \ , \label{bm5}
\eea
are quantised exactly in a way consistent with condition (\ref{bm2}). Notice that the system carries fundamental F1 charges along $x_4$ and $x_5$, 
$\Pi_4 = Q_{F1,4}$,  $\Pi_5 = Q_{F1,5}$.  

\section{S-duality and Magnetic Fields}
\label{s3}
It is illuminating to perform an S-duality of the previous configuration.\footnote{The following arguments were suggested to us by I. Bena.} For that purpose  one adds one extra internal coordinate and considers a D3 brane wrapping $x_0, x_4, x_5, x_6$, with the same
internal electric field.  After S-duality, one gets a D3 brane with the same worldvolume,
with D3 charge $T_3$ and internal magnetic field ${\bf B} = (B_4,B_5,B_6)= (B \cos \beta, B \sin \beta,0) $ or, equivalently, with D1 charges. Indeed, there are two induced D1 charges:
\bea
&&  D1 \ , \ {\rm worldvolume}  \quad x_0, x_4  \qquad  {\rm of \  charge} \quad Q_1^{(4)} = F_{56} = T_1 \frac{E \cos \beta}{\sqrt{1-E^2}} \sim B_4\ , \nonumber \\
&&D1' \ , \ {\rm worldvolume}  \quad x_0, x_5  \qquad {\rm of \ charge}  \quad Q_1^{(5)} = F_{46} =  - T_1 \frac{E \sin \beta}{\sqrt{1-E^2}} \sim B_5 \ . \label{om1}
\eea 
On the other hand, the magnetic fields should satisfy the following (Dirac) quantisation conditions
\begin{align}
B_5 &= B \sin \beta = - F_{46} =\frac{n p}{2 \pi R_4 R_6} \ , &  B_4& = B \cos \beta =  F_{56} = \frac{n q}{2 \pi R_5 R_6}  \ , \label{om2}
\end{align}
where $n$ is an arbitrary integer and $p,q$ are coprime integers. 
From above one obtains the quantisation condition for the angle $\beta$ defining the orientation of the magnetic field (the same by S-duality as the orientation of the
original electric field)
\be
\tan \beta = -\frac{ F_{46}} {F_{56}} =   \frac{p}{q} \frac{R_5}{R_4}   \ . \label{om3}
\ee
 which is again consistent with  (\ref{bm2}) and independent of $R_6$.\\
 \
 It is also useful  to perform a T-duality along $x_6$, turning the D3 brane into a D2 brane which defines a plane in $x_4,x_5,x_6$.  
 The initial electric field is traded for an angle $E = \sin \theta$.  The brane charges are traded into the orientation of the D2 brane plane, defined by the vector ${\bf n}= 
 {\bf Q} = (Q_4,Q_5,Q_6)$  perpendicular to it, where
 \bea
 && Q_{045} = Q_2 = Q \cos \theta \equiv Q_6 \ , \nonumber \\
 && Q_{046} = Q_2  \frac{E \cos \beta}{\sqrt{1-E^2}} = Q \sin \theta \cos \beta \equiv Q_5 \ , \nonumber \\
  && Q_{056} = - Q_2  \frac{E \sin \beta}{\sqrt{1-E^2}} = - Q \sin \theta \sin \beta \equiv Q_4 \  ,  \label{om4}
  \eea
  where $Q_2$ is the standard D2 brane charge and $Q = Q_2/\cos \theta$ is the tension of a rotated brane. From the rotated brane tension one can also identify the T-dual
  magnetic field $B = \tan \theta$. The plane defining the worldvolume of the D2 brane is given by the equation 
 \be
  - \sin \theta \sin \beta \ x_4 + \sin \theta \cos \beta \ x_5 + \cos \theta \ x_6 = 0   \ .  \label{om5} 
 \ee   
The projection of the brane on the $x_4,x_5$ torus  (for $\theta \not=0$) is $- \sin \beta \ x_4 + \cos \beta \ x_5=0$, which is a rotated brane. The length of the brane is finite
 if $\tan \beta = (p R_5)/(q R_4)$. 
       Notice that the S-dual magnetic field is parallel to the original electric field. 
 
 \subsection{Geometrical interpretation}
\label{s31}

 Let us consider a D3 brane on a square 3-torus of coordinates $x,y,z$ and corresponding radii $2 \pi R_{i}$, $i=1,2,3$, with a constant worldvolume magnetic field 
 ${\bf B} = (B_x,B_y,B_z)= (0,B \cos \beta, B \sin \beta) $. The Dirac quantisation condition implies 
 \begin{align}
B_z &= B \sin \beta =  F_{xy} =  \frac{n p_3}{2 \pi R_1 R_2} \ , &  B_y& = B \cos \beta =  - F_{xz} = \frac{n p_2}{2 \pi R_1 R_3}  \ , \label{g1}
\end{align}
where $n$ is an arbitrary integer and $p_2,p_3$ are coprime integers.   The quantisation conditions imply that the direction of the magnetic field and its values are quantised 
\be
 \tan \beta = \frac{p_3 R_3}{p_2 R_2} \ , \quad\quad q B = \frac{n R_\parallel}{2 \pi R_1R_2 R_3}   \ , \label{g2}
  \ee
  where   $R_\parallel = \sqrt{p_2^2 R_2^2 + p_3^2 R_3^2}$ is the length of the coordinate parallel to the magnetic field. Since $(p_2,p_3)$ are (coprime) integers by the Dirac quantisation conditions, the coordinate  $x_{\parallel}$ is therefore compact. Let us denote by $(e_1, e_2, e_3) \in   H_1(\mathbb{T}^3, \mathbb{Z})$ the 1-cycles generating the integral torus homology. It is convenient in what follows to introduce also the basis of one-forms in the cohomology $(\gamma_1,\gamma_2,\gamma_3) \in H^1(\mathbb{T}^3,\mathbb{Z})$, which by (de Rham) duality satisfy
\begin{align}
\int _{e_i} \gamma_j &= \delta_{ij}  \ , & i,j &=1,2,3\ .  
\end{align}
Since the periodicities of the coordinates $x,y,z$ are $2 \pi R_{i}$ then in terms of the coordinate differentials one has  
\begin{align}
\gamma_1 &= \frac{dx}{2 \pi R_1} \ , & \gamma_2 &= \frac{dy}{2 \pi R_2}\ , & \gamma_3 &= \frac{dz}{2 \pi R_3}\ .
\end{align}
Furthermore, we define also the (Poincar\' e) dual two-forms $\{ \beta_i \}_{i=1,2,3}\in H^2(\mathbb{T}^3,\mathbb{Z})$ satisfying
 \be
\int_{\mathbb{T}^3} \gamma_i \wedge \beta_j = \delta_{ij} \ ,
\ee 
and thus, expressed in terms of $\gamma_i$, are given by
\begin{align}
\beta_1 &= \gamma_2 \wedge \gamma_3 \ , & \beta_2 & =-  \gamma_1 \wedge \gamma_3 \ , &\beta_3 &= \gamma_1 \wedge \gamma_2\ .
\end{align}
 In terms of these, one can identify the cycle parallel to the magnetic field ${\bf B}$ and the two-form field strength as 
\be  
 e_{\parallel} = p_2 e_2 + p_3 e_3  \ , \quad\quad \frac{F}{2 \pi} =  n p_2 \beta_2 + n p_3 \beta_3 \ , \label{g3} 
 \ee
 such that the magnetic field $F/(2 \pi)$ is in the integral cohomology $H^2(\mathbb{T}^3,\mathbb{Z})$ of the torus as expected from the Dirac quantisation conditions.   
 
  It is well-known (and easy to check from boundary conditions on the open strings) that a T-duality maps a D3 with a magnetic field into a D2 brane with no worldvolume flux, but rotated. In our case, the T-duality is performed on $x$, whereas the rotated D2 brane is defined by the normal vector  
 \be 
  {\bf n} = (\cos \theta , - \sin \theta \sin \beta , \sin \theta \cos \beta)   \ , \label{g4}  
  \ee
  where the angles $\theta,\beta$ are determined by 
  \be
\tan \theta = \frac{n R_\parallel  R'_1} {m R_2 R_3} \ , \quad\quad \tan \beta = \frac{p_3 R_3}{p_2 R_2}    \ .  \label{g5}
  \ee  
  In (\ref{g5}) $R'_1$ is the T-dual radius $1/2 R_1$, and we added a second `wrapping number' $m$, similar to the one in the first reference in \cite{intersecting}, corresponding to multi-wrapped D-branes with $m$ units of elementary magnetic flux. The integers $(n,m)$ have to be coprime, otherwise one interprets their greatest common divisor as the number of distinct branes. From above one can infer that the D2 brane
  wraps now the following integral 2-cycle 
  \be
  C_2 = m \, e_2 \otimes e_3 + n p_3 \, e'_1 \otimes e_3 + n p_2  \, e'_1 \otimes e_2   \ ,  \label{g6}  
   \ee
where we have introduced the T-dual cycle $e_1'$ of length $2\pi R_1'$. Notice that the product 2-cycles $\{ e'_1 \otimes e_2 ,  e'_1 \otimes e_3, e_2 \otimes e_3 \}$ generate indeed the integral homology $H_2(\mathbb{T}^3, \mathbb{Z})$ (of the dual torus).     
  The tension of the brane is now proportional to its surface
  \be
  T_{\rm D_2} \sim S_{C_2} = (2 \pi)^2  T_2 \sqrt{m^2 R_2^2 R_3^2 + n^2 p_3^2  R_1^{'2} R_3^2 + n^2 p_2^2  R_1^{'2} R_2^3 }  \ .  \label{g7} 
  \ee
  A test of this results is that it is  indeed proportional to the Born-Infeld action of the original D3 brane with worldvolume magnetic field
  \be
  T_{\rm D_3} \sim T_3 \int \sqrt{1 + B^2} \ . \label{g8} 
  \ee  
 
\section{Quantum Mechanics with Magnetic and Electric Fields in Internal Spaces} 
\label{s4}
\subsection{Internal Magnetic Fields}

Let us consider a magnetic field in a three torus $\mathbb{T}^3$ with coordinates $x,y,z$ such that the corresponding vector ${\bf B}=(0,B \cos \beta, B \sin \beta)$ points in an arbitrary direction in the plane $(y,z)$ defined by the angle $\beta$. Then, the non-zero field strength components are the following
\be
F_{xy} = B \sin \beta \ , \quad\quad F_{xz} = - B \cos \beta  \ .  \label{qm1}
\ee
Various gauge choices are possible, but some are more convenient than others for writing down wave functions with appropriate periodicity conditions in the internal space.
We will first make use of the gauge choice
\be
A_x = 0 \ , \quad\quad A_y = B \sin \beta \ x \ , \quad\quad A_z = -B \cos \beta\  x   \ ,  \label{qm2}
\ee
which has the property of being invariant under rotations and translations in the plane $(y,z)$. The gauge above leads to the quantum mechanical charged particle hamiltonian
\be
H \ = \ \frac{1}{2} p_x^2 + \frac{1}{2} (p_y-q B \sin \beta \ x)^2 + \frac{1}{2} (p_z+q B \cos \beta \ x)^2   \ .  \label{qm3}
\ee
The non-zero components $A_y$ and $A_z$ of the potential transform non-trivially only under the torus shifts in the $x$ direction in such a way that the boundary conditions
\be
 A_y (x+2 \pi R_1,y,z) =  A_y (x,y,z) + 2 \pi R_1 B \sin \beta  \ , \quad A_z (x+2 \pi R_1,y,z) =  A_z (x,y,z) - 2 \pi R_1 B \cos \beta \  \label{qm4}
  \ee
correspond to a gauge transformation of parameter $\theta = 2 \pi R_1 B (\sin \beta\, y - \cos \beta\, z)$. The gauge group element  
 \be
 U = e^{i q \theta} = e^{2 \pi i  R_1 qB (\sin \beta \, y - \cos \beta \, z)}  \label{qm4}
  \ee
 is uni-valued on the torus if and only if the components of the magnetic field $F_{xy},F_{xz}$ are quantised as
 \be
 2 \pi R_1 R_2 q B \sin \beta =n\,  p_3 \ , \quad\quad  - 2 \pi R_1 R_3 q B \cos \beta = - n\, p_2  \ , \label{qm5}
   \ee 
 where the integers $p_2,p_3$ are coprime and are identified with the wrapping numbers of the 1-cycle (the lattice vector of minimal length) parallel to ${\bf B}$ in the sub-torus $\mathbb{T}^2_{yz} \in \mathbb{T}^3$. The conditions above are nothing else than the generalised Dirac quantisation conditions (see Section \cite{s31})
 \begin{align}
 \frac{1}{2\pi}\int_{e_1\otimes e_2} F &= n\, p_3 \ , &    \frac{1}{2\pi}\int_{e_1 \otimes e_3} F &= -n \, p_2 \ . 
 \end{align}
 From eq. \eqref{qm5} one can infer that the direction of the magnetic field ${\bf B}$ has to be rational and that its norm, $B$, is quantised. Indeed, we have again the identities 
 \be
 \tan \beta = \frac{p_3 R_3}{p_2 R_2} \ , \quad\quad q B = \frac{n R_\parallel}{2 \pi R_1R_2 R_3}   \ ,  \label{qm6}
  \ee
 where $R_\parallel$ is the periodicity in the direction parallel to ${\bf B}$, given by the length of the corresponding torus cycle, that we denote by $\vec e_\parallel$, having the wrapping numbers $(p_2,p_3)$
 \begin{align}
 \vec e_\parallel &= p_2 \vec e_2 + p_3 \vec e_3  \ , &   R_\parallel &= \sqrt{p_2^2 R_2^2 + p_3^2 R_3^2} \ . 
 \end{align}
 Above, we denote by $\vec e_2$ and $\vec e_3$ the vectors generating the torus lattice $\mathbb{T}^2_{yz}$. It is useful to introduce also the distance between two windings (see Figure \ref{magnetic-image}) of the 1-cycle $\vec e_\parallel \in H_1(\mathbb{T}^2_{yz})$ as follows
 \be
 D_{\parallel} = \frac{R_2 R_3}{R_\parallel}  \ . \label{dw}
 \ee
 The hamiltonian assumes its simplest form in terms of coordinates parallel and perpendicular to the magnetic field vector ${\bf B}$, defined as
 \bea
 && x_{\parallel}  \ = \  \cos \beta \, y +   \sin \beta \, z  \ , \nonumber \\
 && x_{\perp} \ =  \  - \sin \beta \, y +   \cos \beta \, z  \ ,  \label{qm7}
   \eea
  with similar expressions for the parallel and perpendicular momenta.  Notice that the change of coordinates above is not a torus reparametrisation. {\it A priori} the direction defined by $x_\perp$ may not be compact/periodic (though $x_\parallel$ is always compact due to the Dirac quantisation conditions). The new coordinates simplify the interpretation, since the hamiltonian (\ref{qm3}) can be rewritten in the form
 \be
H \ = \ \frac{1}{2} p_x^2 + \frac{1}{2} p^2_\parallel +   \frac{1}{2} (p_\perp + q B x )^2  \ ,  \label{qm03}
\ee
 where it is easy to identify the conserved momenta $p_{\parallel}$ and $ p_\perp$ (or equivalently $p_y, p_z$), the center of mass of the harmonic oscillator $x_{cm} = p_{\perp}/qB$ and the Larmor frequency $\omega_L=q B$. Thus, each energy eigenvalue is determined by the harmonic oscillator level $\lambda$ and by the parallel momentum $p_\parallel$. The resulting wave function in the non-compact space ($\mathbb{R}^3$) has the form
 \be
 \Psi(x,y,z) = e^{i p_y y}e^{i p_z z} \psi_\lambda \left(x- \frac{\sin \beta \, p_y - \cos \beta\, p_z}{qB}  \right)  \  , 
  \ee
where $\psi_\lambda$ is the harmonic oscillator eigenfunction of level $\lambda$. In the chosen gauge \eqref{qm2}, the translation operators in the three directions of the torus result
  \bea
  && U_x = e^{2 \pi i R_1  (P_x - q B y \sin \beta + q B z \cos \beta )} \ , \nonumber \\
  && U_y = e^{2 \pi i  R_2 P_y}  \ , \ \quad  U_z = e^{2 \pi i  R_3 P_z}  \ , \label{qm9}  
     \eea  
leading to the following periodicity conditions
 \bea
 && \Psi (x + 2 \pi R_1, y,z) = e^{2 \pi i R_1  q B (y \sin \beta - z \cos \beta) }  \ \Psi (x, y,z)   \ , \nonumber \\
 &&  \Psi (x, y + 2 \pi R_2,z) =  \Psi (x, y,z)  \ , \quad\quad \Psi (x,y,z + 2 \pi R_3) =  
 \Psi (x, y,z)    \ , \label{qm10}
  \eea 
necessary for the wave function to be well defined on the torus $\mathbb{T}^3$.  Equivalently, the periodicities above follow also from the bundle transition function in eq. \eqref{qm4}, combined with the fact that the gauge potential \eqref{qm2} is invariant under translations in $y$ and $z$. The periodicity of the wave function in the coordinates $y,z$ imply that we have the standard Kaluza-Klein (KK) quantisations for the momenta
\begin{equation}
p_y = \frac{q_2}{R_2} \ , \quad\quad  p_z = \frac{q_3}{R_3} \ . 
\end{equation}       
In order to build a wave function which respects also the quasi-periodicity in $x$ one can superpose $(p_y,p_z)$, or equivalently the momentum numbers $(q_2,q_3)$, such that the energy level remains fixed (and thus also $p_\parallel$). Indeed, a generic torus shift in the $x$ direction generates an image (of the harmonic oscillator wave function) with momenta of the form
\be
x\rightarrow x+ 2 \pi m R_1\quad  \implies  \quad 
\left \{ 
\begin{split}
&p_y \rightarrow p_y - 2 \pi m qBR_1 \sin \beta\\
&p_z \rightarrow p_z+2 \pi m qB R_1 \cos \beta
\end{split} \right.
\ee
where $m$ indexes the images. Taking into account the quantisation of the angle $\beta$ and of the magnetic field $B$ one can then see that, on the torus, there is the following equivalence relation
\be
(q_2,q_3) \sim (q_2 - m n p_3, q_3 + m n p_2)
\ee 
which leaves $p_\parallel$ invariant. The solution for the wave functions with correct periodicities can be found by summing over all the images
 \be
 \Psi (x, y,z) = \sum_{m\in \mathbb{Z}} e^{i (p_y - m q B R_1 \sin \beta) y + i (p_z + m q B R_1 \cos \beta) z  }  \ \Psi_{\lambda} \left (x + 2 \pi m R_1 + \frac{-\sin \beta \ p_y + \cos \beta \ p_z}{q B} \right) 
  \  . \label{qm11} 
   \ee 
Notice that the term with $m=0$ corresponds to the wave function in non-compact space. In order to understand the quantisation of $p_\parallel$ and the Landau level degeneracy it is convenient to introduce a torus reparametrisation, i.e. an $SL(2,\mathbb{Z})$ matrix $M$, such that one of the basis vectors of the lattice  is given by $\vec e_\parallel$. We can then write
  \be
\left(  \begin{array}{c}
    \vec e_\parallel \\ 
    \vec e_* \\ 
  \end{array} \right) =  \left(  \begin{array}{cc}
    p_2 & p_3 \\ 
    l_2 & l_3 \\ 
  \end{array} \right)  \left(  \begin{array}{c}
    \vec e_2 \\ 
    \vec e_3 \\ 
  \end{array} \right)  \ . 
\ee 
It should be observed that the vector $\vec e_*$ is in general not orthogonal to $\vec e_\parallel$. Only in the limits $\beta = 0, \pi/2$ we can choose $\vec e_*$ to be either $\vec e_3$ or $- \vec e_2$ (and thus orthogonal to $\vec e_\parallel$ that, in this case,  becomes either $\vec e_2$ or $\vec e_3$). We have also the identities
 \begin{align}
 M^{-1}&=  \left(  \begin{array}{cc}
    l_3 & -p_3 \\ 
    -l_2 & p_2 \\ 
  \end{array} \right) \ , &&  & \det M &=\det M^{-1} = p_2 l_3 - p_3 l_2 = 1 \ . 
 \end{align}
Taking into account our conventions for the periodicity of the coordinates, namely $y \sim y+ 2 \pi R_2$ and $z \sim z+ 2 \pi R_3$,  the momentum quantum numbers transform, under the above reparametrisation, as
\be
\left(  \begin{array}{c}
    q_2 \\ 
    q_3 \\ 
  \end{array} \right) =  \left(  \begin{array}{cc}
    l_3 & -p_3 \\ 
    -l_2 & p_2 \\ 
  \end{array} \right)  \left(  \begin{array}{c}
    k \\ 
    j \\ 
  \end{array} \right)  \ , 
\label{qqkj}
\ee 
 where $k$ is associated to the parallel direction and $j$ is associated to the $*$ direction.  Making use of \eqref{qqkj} and of the quantisation of the angle $\beta$, one can easily show that the parallel and perpendicular momenta are quantised as
 \begin{align}
 p_\parallel &= \frac{k}{R_\parallel}(p_2 l_3 - p_3 l_2)= \frac{k}{R_\parallel} \ , \label{wrq}\\
 p_\perp & = \frac{1}{R_2 R_3} \left [j R_\parallel - \frac{k}{R_\parallel}\left (p_2 l_2 R_2^2 + p_3 l_3 R_3^2 \right) \right] \ . 
 \end{align}
 Taking into account the quantisation of $B$ one obtains that the center of mass of the oscillator (for the wave function on the torus) has the expression
 \be
 x_{cm} = \frac{p_\perp}{qB}+2 \pi m R_1 = \left[\frac{j+mn}{n} - \frac{k}{n R_\parallel^2}(p_2 l_2 R_2^2 + p_3 l_3 R_3^2) \right] 2 \pi R_1 \ . 
 \ee
For a given energy level, that is for fixed $\lambda$ and $k$, the center of mass takes a finite number $n$ of discrete values $j=0,1,\dots, n-1$, corresponding exactly to the Landau level degeneracy. It should be stressed that in the zero angle limit one takes $(p_2,p_3)= (1,0)$ and $(l_2,l_3) = (0,1)$, so that the second term in the equation above depending on $k$ disappears, thus recovering the known result of $x_{cm} =2\pi R_1  (j+mn)  /n $. The degeneracy can be checked also from the point of view of the annulus amplitude of open strings in the presence of the considered magnetic field. After taking into account the quantisation of the parallel momentum one has\footnote{Our partition functions in magnetic fields should be multiplied by $1/(2 \pi)^D$, where $D$ is the number of non-compact dimensions. According to our standard conventions \cite{orientifolds}, we will not write explicitly this factor.}
\begin{align}
\mathcal{A} =\frac{V_{7} V_3}2\, qB \frac{1}{R_\parallel}\int_0^\infty \frac{d\tau_2}{\tau_2^{9/2}}\sum_{\alpha,\beta}  c_{\alpha \beta} \frac{\vartheta \left[ \alpha \atop \beta\right] \left( \epsilon \frac{i\tau_2}{2}\Big |  \frac{i \tau_2}{2}\right)}{\vartheta \left[\alpha \atop \beta\right] \left ( 0 \Big | \frac{i\tau_2}{2}\right)} \frac{\vartheta^4\left[\alpha \atop \beta \right] \left(0 \Big | \frac{i\tau_2}{2}\right)}{\eta^{12} \left(\frac{i\tau_2}{2}\right)}\frac{i \eta^3\left(\frac{i\tau_2}{2}\right)}{\vartheta_1\left ( \epsilon \tfrac{i\tau_2}{2}|\tfrac{i\tau_2}{2}\right)} \sum_{k}e^{- \pi \tau_2 \frac{k^2}{2 R_\parallel^2}}
 \label{magann}\end{align}
Since the identity $V_3 qB/R_\parallel = n$ holds, the amplitude above correctly counts the particle propagation with degeneracy $n$. The tree-level closed (transverse) string amplitude is then given by
\begin{align}
  \mathcal{\tilde A} &=2^{-5}\frac{V_{7}V_3}2\, qB \,  \int_0^\infty dl \sum_{\alpha,\beta}  c_{\alpha \beta} \frac{\vartheta \left[ \alpha \atop \beta\right] \left( \epsilon  |  il \right)}{\vartheta \left[\alpha \atop \beta\right] \left ( 0 | il \right)} \frac{\vartheta^4\left[\alpha \atop \beta \right] \left(0  | il \right)}{\eta^{12} \left(i l\right)}\frac{i \eta^3(il)}{\vartheta_1\left (\epsilon |il\right)}\, \tilde P_{\tilde k} \ , 
\end{align}
where the dual lattice sum has the form
\be
\tilde P_{\tilde k}  = \sum_{\tilde k} e^{- \pi l \tilde k^2 R_\parallel^2} \ . 
\ee
It correctly describes the coupling of the branes to the (closed-string) winding  states and the Born-Infeld structure of the magnetised brane tension. Indeed, by using the identity $\sin \pi \epsilon = qB/\sqrt{1+q^2 B^2}$ (this factor is contained in $\vartheta_1(\epsilon|il)$) one finds
\be
\mathcal{\tilde A}_c \sim \sqrt{1+q^2 B^2} \ . 
\ee
Another natural choice to study the magnetic field (\ref{qm1})  is the Landau gauge given by the following expression
\bea
&& A_x = - B (\sin \beta \ y - \cos \beta \ z )  \ , \nonumber \\ 
&& A_y= A_z= 0  \ . \label{qm12}
\eea
 It is interesting to notice that the gauge transformation passing from the gauge (\ref{qm12}) to the previous one  (\ref{qm2}), defined by the gauge parameter
 $\theta = - B x (\sin \beta y - \cos \beta z)$,  cannot be well-defined on the whole torus. The charged particle hamiltonian in the gauge of eq. \eqref{qm12} is 
 \be
 H = \frac12 \left[p_x + q B (\sin \beta\, y - \cos \beta\, z) \right]^2 + \frac12 \left(p_y^2 + p_z^2 \right)  \ . 
 \ee
 The potential transforms non-trivially under the torus periodicities in the directions $y$ and $z$, yielding the boundary conditions
 \begin{align}
 A_x(x,y+2\pi R_2, z) &= A_x(x,y,z)-2 \pi R_2 B \sin \beta \ , \nonumber\\
 A_x(x,y, z+2\pi R_3) &= A_x(x,y,z)+2 \pi R_3 B \cos \beta \ . \label{qm14}
 \end{align}
 The torus shifts in the directions $y$ and $z$ generate two gauge transformations with parameters $\theta_1 = - 2\pi R_2 B \sin \beta \, x$ and $\theta_2 =  2\pi R_3 B \cos \beta \, x$. In order for the potential to be well defined on the torus one needs the following transition functions  
 \begin{align}
 U_1 &= e^{i q \theta_1} = e^{-2\pi i R_2 qB \sin \beta \, x} \quad , &  U_2 &= e^{i q \theta_2} = e^{2\pi i R_3 qB \cos \beta \, x} \label{qm}
 \end{align}
 to be single valued on the torus.  It is easy to see that this holds true if and only if the quantisation conditions of the projections of the magnetic field in eqs. \eqref{qm5} are satisfied. In terms of parallel and perpendicular positions (and momenta) defined in eq. \eqref{qm7} one can write the hamiltonian as follows
 \be
 H=\frac12 p_\perp^2 + \frac12 q^2 B^2 \left(x_\perp - \frac{p_x}{qB} \right)^2 + \frac12 p_\parallel^2 \ , 
 \ee
where again one can identify the center of mass position of the harmonic oscillator $x_{cm} = p_x/qB$. Since $p_\parallel$ and $p_x$ commute with the hamiltonian the wave function in a non-compact space is now of the form
\be
\Psi(x,y,z) = e^{ip_x x}e^{i p_\parallel x_\parallel} \Psi_\lambda\left(x_\perp - \frac{p_x}{qB} \right) \ . 
\ee
Making use of eq. \eqref{qm} one finds that the wave function on the three torus has to satisfy the following periodicity conditions
\begin{align}
\Psi (x+2 \pi R_1, y,z) &= \Psi (x,y,z) \ , & \Psi(x,y+2\pi R_2,z) = e^{-2\pi i R_2 q B \sin \beta\, x} \Psi(x,y,z)\ ,\nonumber\\
\Psi(x,y,z+2\pi R_3) &= e^{2\pi i R_3 q B \cos \beta\, x} \Psi(x,y,z) \ . 
\end{align}
 Since in the direction $x$ the solution is a plane wave, it follows that the momentum $p_x$ is quantised as $p_x = m /R_1$. In order to build a wave function with the correct (quasi)-periodicities one can use
 \be
 \Psi_j(x,y,z) = \sum_{m\in \mathbb{Z}} e^{ i (j+m n)\frac{x}{R_1}} e^{i (\cos \beta \, p_y + \sin \beta \, p_z)(\cos \beta \, y + \sin \beta \, z)} \Psi_\lambda \left(-\sin \beta \, y + \cos \beta \, z -  \frac{j+mn}{n}2 \pi D_\parallel \right)
 \ee
where $D_\parallel$ was defined in eq. \eqref{dw}. It is easy to see that one has the correct quasi-periodicities, after a redefinition of the summation index m of the form
\begin{align}
y & \rightarrow y+2\pi R_2\ ,  &&  & m &\rightarrow m' = m + 2 \pi R_1 R_2 q B \sin \beta \ , \nonumber\\
z & \rightarrow z+2\pi R_3 \ , &&  & m &\rightarrow m' =m - 2 \pi R_1 R_3 q B \cos \beta \ . 
\end{align}
The shift of the summation index has to be an integer! Indeed this is assured by the quantisation conditions of the magnetic field components in eq. \eqref{qm5}. In addition,
it seems that we must also satisfy the following quantisation conditions for the momenta
\begin{align}
2 \pi R_2 (\cos \beta \, p_y + \sin \beta \ p_z)\cos \beta &= 2 \pi k\, k_3 \ , \nonumber\\
2 \pi R_3 (\cos \beta \, p_y + \sin \beta p_z)\sin \beta & = - 2 \pi k \,k_2  \ , \label{momq}
\end{align}
where $k_2$, $k_3$ are coprime integers that can be identified with the wrapping numbers on the torus 1-cycle $\vec e_\perp$ orthogonal to the magnetic field. From the ratio of eqs. \eqref{momq} we can readily extract a quantisation condition for the $\beta$ angle of the form
 \begin{align}
 \tan \beta =  -\frac{k_2 R_2}{k_3 R_3} = - \cot \left(\beta + \frac{\pi}{2}\right) \ . \label{ang2}
 \end{align}
Moreover, since we obtain that the angle $\beta + \pi/2$ is rational, then also the coordinate $x_\perp$ appear to be compact with an effective radius given by
\be
R_\perp = \sqrt{k_2^2 R_2^2 + k_3^2 R_3^2} \ . 
\ee
If both the parallel and orthogonal directions are required to be compact, we can then write the following  $GL(2,\mathbb{Z})$ transformation
  \be
\left(  \begin{array}{c}
    \vec e_\parallel \\ 
    \vec e_\perp \\ 
  \end{array} \right) =  \left(  \begin{array}{cc}
    p_2 & p_3 \\ 
    k_2 & k_3 \\ 
  \end{array} \right)  \left(  \begin{array}{c}
    \vec e_2 \\ 
    \vec e_3 \\ 
  \end{array} \right)  \ , 
\ee 
that we denote by $\tilde M$. Notice that we have the identitites 
\be
\det \tilde M = p_2k_3 - p_3 k_2  = \frac{R_\parallel R_\perp}{R_2 R_3} \ . 
\ee
Geometrically, the ratio of the volumes is thus given by the intersection number of the
parallel cycle with the orthogonal one (this can  be easily shown by using $\vec e_\parallel \wedge \vec e_\perp = (\det \tilde M) \vec e_2 \wedge \vec e_3$).
\begin{figure}[h]
\begin{center}
\includegraphics[width=\textwidth]{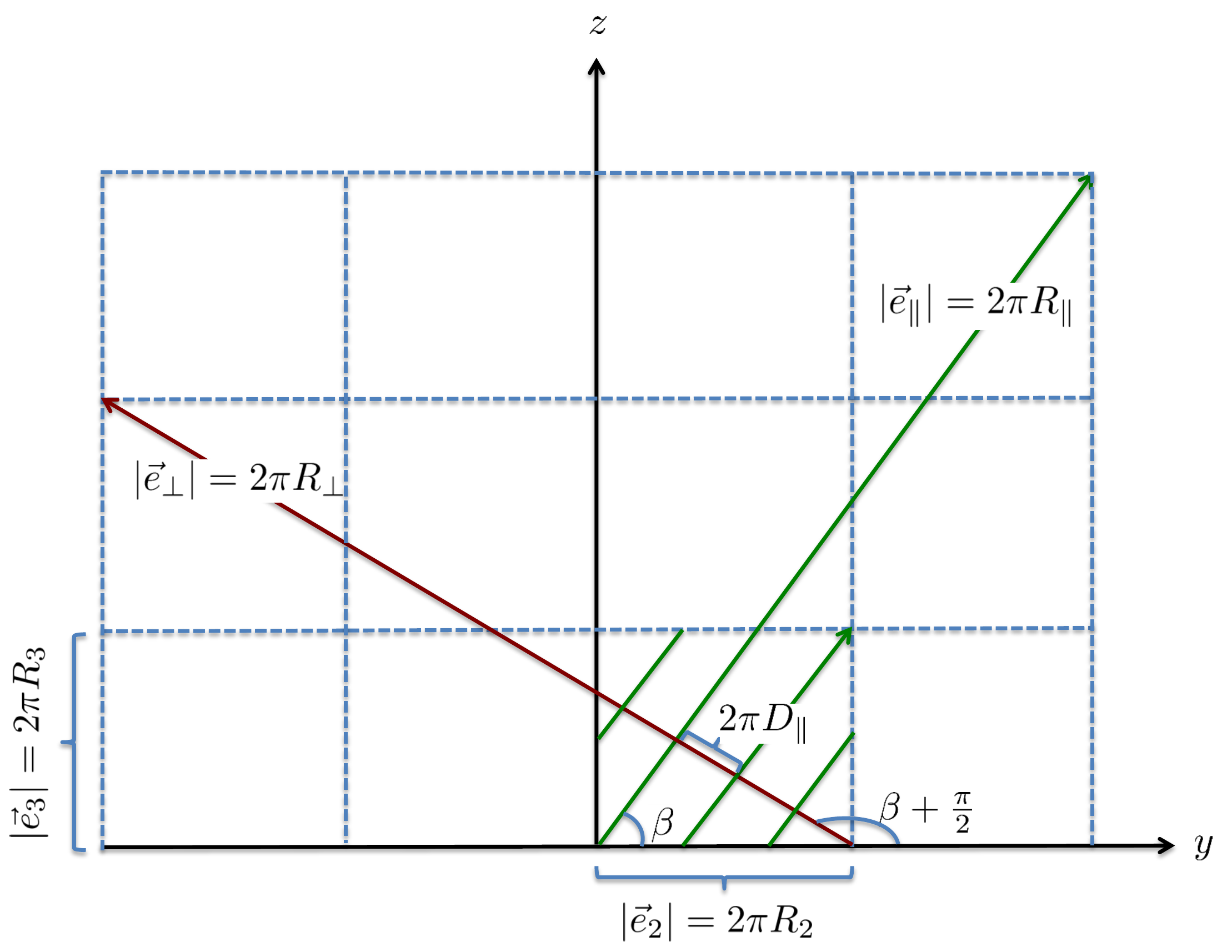}
\end{center}
\caption{We depict the lattice vectors $\vec e_\parallel$ with wrapping numbers $(2,3)$ and $\vec e_\perp$ with wrapping numbers $(-3,2)$. The distance bewtween two winding of the parallel cycle $D_\parallel$ is also illustrated. From eq. \eqref{rational} the complex structure of the torus has to be in such a way as to have $R_2 = R_3$.}\label{magnetic-image}
\end{figure}
From eq. \eqref{momq} one can extract a quantisation condition for the momentum $p_\parallel$ of the form
\be
p_\parallel = \frac{k R_\perp}{R_2 R_3} =\frac{k}{R_\parallel} \det \tilde M \ . 
\ee
The correct normalization requires a ratio with the determinant of $\tilde M \in GL(2,\mathbb{Z})$. Thus, one finally gets 
\be
p_\parallel = \frac{k}{R_\parallel} \ ,
\ee
as in eq. \eqref{wrq}. Combining the eq. \eqref{ang2} with the condition for the angle in \eqref{qm6}, we obtain that the ratio of the squares of the radii $R_2$ and $R_3$ is fixed to be a rational number
 \be
 \frac{R_3^2}{R_2^2} = - \frac{k_2 p_2 }{ k_3 p_3} \in \mathbb{Q} \ . \label{rational}
 \ee
Notice that the identity \eqref{rational} can be obtained by imposing that $\vec e_\parallel$ and $\vec e_\perp$ are orthogonal to one another. In conclusion, in this gauge we obtain that both the parallel coordinate  $x_\parallel$ and the perpendicular coordinate  $x_\perp$ have to be compact and, consequently, the squared modulus  of the complex structure of the sub-torus $\mathbb{T}^2_{yz}$ is fixed to be a rational number determined by the wrappings of the parallel and orthogonal (with respect to the vector magnetic field ${\bf B}$) 1-cycles. It should be possible and  interesting to  construct a more general wave function which does not require the condition \eqref{rational}. Indeed, from the point of view of the string cylinder amplitude in eq. \eqref{magann} one does not need the orthogonal coordinate to be compact in order to correctly interpret the amplitude. If, in addition to the generalised Dirac quantisation conditions, one needs to impose also the constraint \eqref{rational}, then this result could be very important for the stabilisation of complex structure moduli fields in string theory.  

\subsection{Internal Electric Fields}

The action of S-duality is in such a way that the electric and magnetic fields are parallel. As we have seen, Dirac quantisation conditions imply that the coordinates parallel to the magnetic field is compact. In turn, making use of S-duality, one obtains that the direction of the electric field is also quantised (rational from the point of view of the torus). 
Let us consider an electric field in an arbitrary direction in the plane $yz$, ${\bf E} = (E_y,E_z)=(E \cos \beta, E \sin \beta)$.  Our conventions for the electric field are such that ${\bf E}$ is parallel to the magnetic field considered in the previous section. This is convenient, as many formulas from the magnetic field side translate to the electric field case simply by analytic continuation ($x \rightarrow i x_0, p_x \rightarrow i p_0, B \rightarrow -i E$). 

In what follows we use in a slightly abusive  way the language of hamiltonian and wave function. Strictly speaking, there are no stationary states and all the hamiltonians
below are not hamiltonians in the usual sense: either they are time-dependent with inverted harmonic oscillators or they involve oscillators with a center of mass determined by the energy. In fact, particles in an electric field are accelerated and radiate. The results below are better understood actually in terms of analytic continuation from the magnetic field case, and what we call hamiltonians will correspond actually to the zero-mode part of the string hamiltonians in later sections. 
The analytic continuation leads to a (complex)  partition function which does not have the standard interpretation from quantum mechanics or string theory. It contains, however, physical information in the sense of encoding the energy loss by charged D-branes in the presence of an electric field, that is the natural generalization of the Schwinger pair production in quantum field theory \cite{schwinger}.  

We start again with the gauge where the wave function is manifestly periodic in
$y$ and $z$, namely 
\be
A_0 = 0 \ , \quad \quad A_y = E \cos \beta \ x_0 \ , \quad\quad A_z = E \sin \beta\  x_0   \ ,  \label{qm13}
\ee
which leads to the quantum mechanical hamiltonian
\be
H \ = \ - \frac{1}{2} p_0^2 + \frac{1}{2} (p_y-q E \cos \beta \ x_0)^2 + \frac{1}{2} (p_z-q E \sin \beta \ x_0)^2   \ .  \label{qm14}
\ee
In terms of the parallel and the orthogonal coordinates, the (time-dependent) hamiltonian is given by
  \be
H \ = \ - \frac{1}{2} p_0^2 + \frac{1}{2} (\sin \beta\,  p_y - \cos \beta\, p_z )^2 +   \frac{1}{2} (\cos \beta\, p_y + \sin \beta\, p_z - q E x_0 )^2  \ ,  \label{qm15}
\ee
 which identifies the conserved momentum $p_{\perp} = - \sin \beta\, p_y +  \cos \beta\, p_z$ and the `center of mass' of the oscillator $x_{cm} = p_{\parallel}/qE$. However, it should be noticed that the above hamiltonian describes an inverted harmonic oscillator. We will treat the system as an analytic continuation of the harmonic oscillator and thus the wave functions that we find are not really physical. Nevertheless, one can use this formal procedure to extract physically relevant results.
 
In this gauge, the translation operators in the directions of the torus are the usual ones
  \be
   U_y = e^{2 \pi i R_2 P_y } \ , \quad \quad  U_z = e^{2 \pi i  R_3 P_z}   \ . \label{qm16}  
  \ee  
As a consequence, since the potential is invariant under translations in the $y$ and $z$ directions, the wave functions have to be periodic
 \be
  \Psi (x_0,y + 2 \pi R_2, z) =    \Psi (x_0,y, z)   \ , \quad\quad   \Psi (x_0,y, z + 2 \pi R_3) =  \Psi (x_0,y, z)   \ . \label{qm17}
  \ee 
From eq. \eqref{qm13}, we obtain that time translations $x_0 \rightarrow x_0+t$ generate a gauge transformation of the potential given by 
\begin{align}
\theta_t(y,z) = E(\cos \beta\, y + \sin \beta\, z)\, t \ .
\end{align}   
Hence, in this gauge, time translations are implemented by gauge transformations. We find that, in addition to the periodicity conditions in eq. \eqref{qm17}, one has to also impose 
\begin{align}
\Psi(x_0+t,y,z) &= e^{i q E (\cos \beta\, y + \sin \beta \, z)\, t}\,\Psi(x_0,y,z)\, , && \text{for all}\ t\in \mathbb{R} \ .
\end{align}      
 The solution for the wave functions periodic on the two torus with the correct implementation of time translations is then
 \be
 \Psi (x_0, y,z) =  \int_{-\infty}^{+\infty} d \alpha\,  e^{i (p_\parallel + \alpha q E)x_\parallel  }\, e^{i p_\perp x_\perp}  \ \Psi_{\lambda} \left(x_0  - \alpha - \frac{p_\parallel}{q E}\right) 
  \ , \label{qm18} 
   \ee 
 with standard KK momenta $p_y = q_2/R_2$, $p_z= q_3/R_3$, and $q_i$ integers. Notice that one can set the parallel momentum $p_\parallel$ to any value by a shift of the integration variable $\alpha$. Formally, it plays the same role as the Landau level degeneracy in the magnetic field case.  If one sets $p_\parallel = 0$ as it is required by the open string boundary conditions, then a quantisation condition for the angle $\beta$ arises as
 \be
 \tan \beta = - \frac{q_2 R_3}{q_3 R_2} \ . \label{e1}
 \ee
 \\
 The identity above implies that the coordinate $x_\parallel$ is compact. However, from a quantum mechanical point of view, neither this choice nor  
 (\ref{e1}) seem to be necessary.  We know, however, from the arguments in Section \ref{s2} that (\ref{e1}) is true. Let us use the following ansatz for the momentum numbers $q_2$ and $q_3$
 \begin{equation}
q_2 = - j\, p_3 \ , \quad\quad q_3  = j\, p_2 \ , \label{ansatz}
  \end{equation}
 where the coprime integers $(p_2,p_3)$ determine the wrapping numbers of the 1-cycle parallel to the electric field. Then the angle quantisation becomes
 \be
 \tan \beta = \frac{p_3 R_3}{p_2 R_2} \ . 
 \ee
Using the Ansatz \eqref{ansatz}, the perpendicular and parallel momenta become
 \begin{equation}
 p_\perp = \frac{j R_\parallel}{R_2 R_3}\ , \quad \quad    p_\parallel = 0 \ .
 \end{equation}
Generically, one expects a quantisation of the perpendicular momentum of the form
\be
p_\perp = \frac{j}{R_\perp} \ , 
\ee
with the orthogonal radius being possibly infinite. As in the magnetic field case, we run again into the rescaling by the factor $R_\parallel R_\perp/R_2 R_3$.
\begin{figure}[h]
\begin{center}
\includegraphics[width=\textwidth]{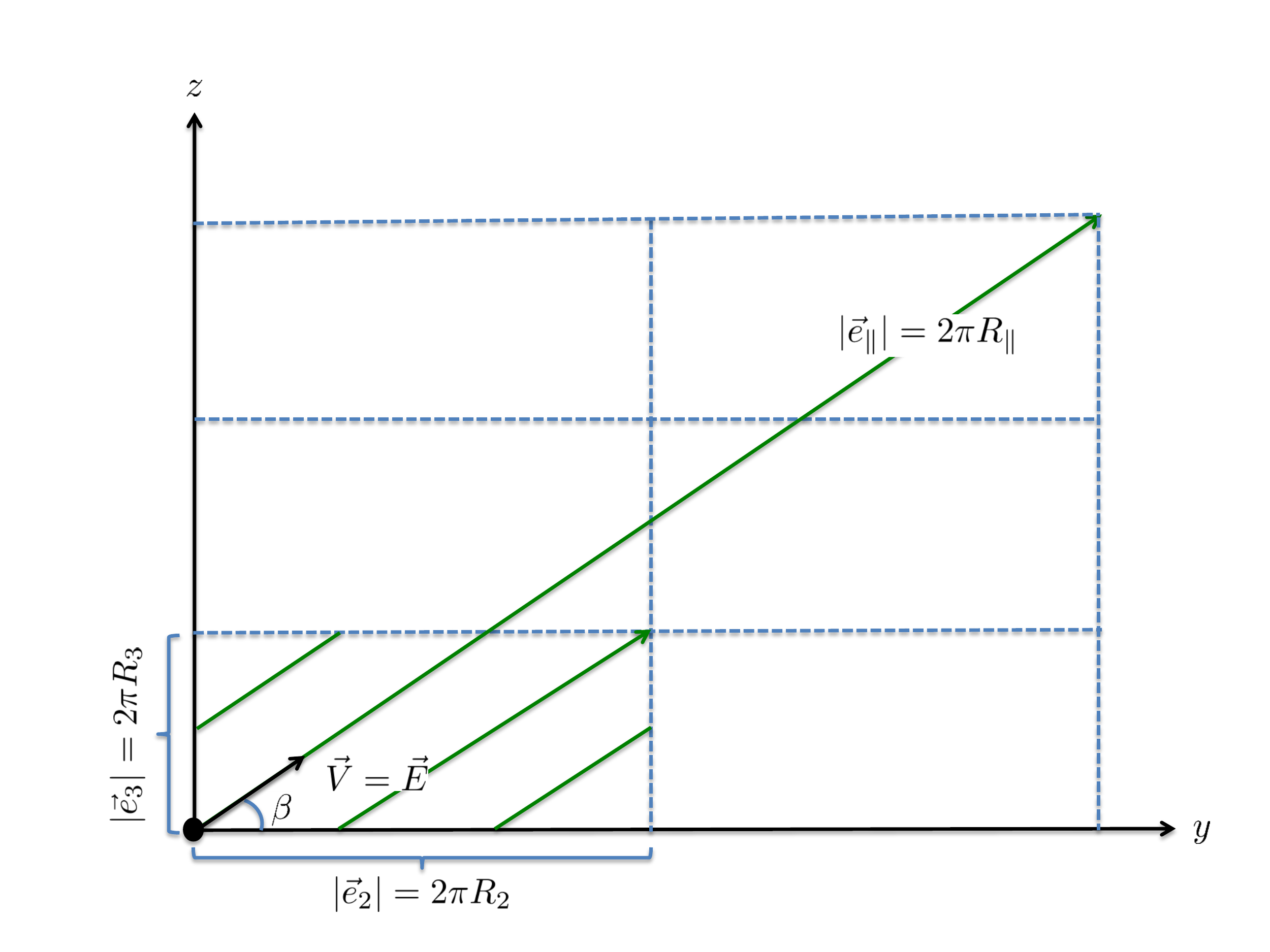}
\end{center}
\caption{We depict a particle moving with velocity $\vec V = \vec E$ corresponding to the T-dual picture of an electric field. Due to the quantisation condition of the angle, the trajectory of the particle inside the two torus is periodic.}\label{magnetic-image}
\end{figure}

Analogously to the case of the magnetic field, we can also consider the gauge corresponding to the Landau one
\bea
&& A_0 = - E (\cos \beta \, y + \sin \beta \, z )  \ , \nonumber \\ 
&& A_y= A_z= 0  \ . \label{qm19}
\eea
Again, the transformation to pass from the gauge (\ref{qm19}) to the previous (\ref{qm13}) is given by the function 
 $\theta = - E x_0 (\cos \beta y + \sin \beta z)$ that,  as before, it is not well-defined on the torus. The hamiltonian can be written as
 \begin{align}
 H &= -\frac12 \left[ p_0 + q E (\cos \beta \, y + \sin \beta \, z)\right]^2 + \frac12 \left(p_y^2 + p_z^2 \right)= \frac12 p_\parallel^2 - \frac{q^2 E^2}{2} \left(x_\parallel + \frac{p_0}{qE} \right)^2  + \frac12 p_\perp^2 \ , 
 \end{align}
 where we make use again of parallel and perpendicular coordinates/momenta. Since the momenta $p_0$ and $p_\perp$ commute with the hamiltonian, the wave function in non-compact space has the form
 \be
 \Psi (x_0, y,z) = e^{ip_0 x_0}e^{i p_\perp x_\perp} \Psi_\lambda \left(x_\parallel + \frac{p_0}{qE} \right) \ . 
 \ee
On the two torus $\mathbb{T}^2$ with periodicities in the $y,z$ coordinates, the wave function has to transform in the following way
\begin{align}
\Psi (x_0, y+ 2 \pi R_2, z) &= e^{-2 \pi i R_2 q E x_0 \cos \beta}\Psi(x_0,y,z) \ , \nonumber \\
 \Psi (x_0, y, z+ 2 \pi R_3) &= e^{-2 \pi i R_3 q E x_0 \sin \beta}\Psi(x_0,y,z) \ , 
\end{align}
as indicated by the transformation of the gauge potential \eqref{qm19} under the same shifts. A wave function with the correct quasi-periodicities can be written as
\begin{align}
&\Psi(x_0,y,z) = \nonumber\\
&\int_{-\infty}^{+\infty} d \alpha \ e^{i (p_0 - \alpha q E) x_0} e^{i (-\sin \beta \, p_y + \cos \beta \, p_z) (- \sin \beta \, y + \cos \beta \, z)} \psi_\lambda \left(\cos \beta \, y + \sin \beta \, z + \alpha+\frac{p_0}{q E}\right).
\end{align}
Indeed, torus lattice shifts correspond to changes of variable in the integral 
\begin{align}
y&\rightarrow y+ 2 \pi R_2 \ , & \alpha &\rightarrow \alpha' = \alpha + 2\pi q E R_2 \cos \beta  \ ,   \nonumber\\
z&\rightarrow y+ 2 \pi R_3 \ , & \alpha &\rightarrow \alpha' = \alpha + 2\pi q E R_3 \sin \beta  \ . 
\end{align} 
Notice that $p_0$ can be set to any value by a shift of the integration variable $\alpha$. In addition, for the wave function to be well defined on the torus, the following quantisation conditions for the momenta seem to be needed
\begin{align}
- 2 \pi R_2 (- \sin \beta \, p_y + \cos \beta \, p_z) \sin \beta &= -2 \pi  j\, p_3  \ , \label{mom1}\\
2 \pi R_3 (- \sin \beta \, p_y + \cos \beta \, p_z) \cos \beta &= 2 \pi j\, p_2  \ . \label{mom2}
\end{align} 
Thus, we see that also in the case of the electric field one would obtain a quantisation condition for the angle $\beta$
 \be
 \tan \beta = \frac{p_3 R_3}{p_2 R_2} \ ,  \label{el2}
 \ee
 which would imply that the coordinate $x_\parallel$ is compact with radius $R_\parallel = \sqrt{p_2^2 R_2^2 + p_3^2 R_3^2}$. Since in the first gauge \eqref{qm13} this quantisation is not really manifest, we consider the condition (\ref{el2}) to be inconclusive from the viewpoint of quantum mechanics, in a similar way to the compactness of the coordinate perpendicular to the magnetic field in the Landau gauge \eqref{qm12} considered in the previous section. However, unlike the latter, the quantisation (\ref{el2})  is predicted by the non-perturbative arguments of Sections \ref{s2} and \ref{s3}. 
Notice that, due to the fact that the coordinate $x_0$ is not compact, we do not have quantisation conditions for the components of the electric field. As a consequence, from a quantum mechanical point of view, we find that the coordinate $x_\perp$ does not have to be compact. Relevant for constructing the CFTs of strings with background electric fields is the fact that the perpendicular momentum is quantised from eqs. \eqref{mom1}-\eqref{mom2} as
 \be
 p_\perp = \frac{j R_\parallel}{R_2 R_3}
 \ee
However, when the coordinate $x_\perp$ is compact, one expects a quantisation of the form $p_\perp = j/R_\perp$, which can be obtained after rescaling with the ratio of the volumes. 
 
To summarise, the quantisation of the direction parallel to the electric field predicted non-perturbatively from the arguments of Sections \ref{s2} and \ref{s3} is not completely manifest from quantum mechanical arguments. String quantisation and one-loop amplitudes in later sections seem also to be consistent with any value of the angle. This implies that the quantisation of $x_\parallel$, whereas unambiguously predicted, is probably a genuine non-perturbative effect, invisible in perturbation theory.
   
\section{Open Strings with Boundary Electric Fields}
\label{s5}
Open strings with (generically different) boundary electromagnetic fields can be described by the following $\sigma$-model action
\begin{align}
S= - \frac{1}{2\pi} \int_{-\infty}^{+\infty}d\tau \int_{0}^\pi d\sigma\, \partial_\alpha X^\mu \partial^\alpha X_\mu-q_1\int_{-\infty}^{+\infty}d\tau \, A_1{}_\mu\, \partial_\tau X^\mu-q_2\int_{-\infty}^{+\infty}d\tau \, A_2{}_\mu\, \partial_\tau X^\mu \ , \label{action}
\end{align}
where the string worldsheet has been taken to be the infinite strip and units are chosen such that
\be
2 \alpha' = 1 \ . 
\ee
 The gauge potentials $A_1$ and $A_2$ are different when considering strings stretched between different stacks of branes. A convenient gauge choice for constant electromagnetic fields is
\begin{align}
A_i^\mu &= - \frac12 F_i^\mu{}_\nu X^\nu\, ,  & i&=1,2 \ .\label{gauge}
\end{align}
Our convention for the boundary charges $q_1$ and $q_2$ is such that the case of neutral (dipole) strings corresponds to the condition
\be
q_1 + q_2 = 0 \ . \label{dipole}
\ee
In the following, for convenience  we shall absorb the charges $q_i$ into redefined field strengths $F_i$
\be
\pi q_i F_i \rightarrow F_i \ . 
\ee
With our choice of gauge in eq. \eqref{gauge} the classical system amounts to the wave equation for the bosonic coordinates $X^\mu(\tau, \sigma)$ together with general (i.e. a combination of Neumann and Dirichlet) boundary conditions
\begin{align}
 \partial_\sigma X^\mu&= F_1^\mu{}_\nu\, \partial_\tau X^\nu\, , & \sigma & = 0 \ , \label{bc1}\\
 \partial_\sigma X^\mu&= - F_2^\mu{}_\nu\, \partial_\tau X^\nu\, , & \sigma & = \pi \ . \label{bc2}
\end{align}
We shall restrict our discussion to electric fields $E$ and $\tilde E$ in the plane $(X^8,X^9)$ and thus the field strengths are of the following form
\begin{align}
F_1& := F_1^\mu{}_\nu = \left(  \begin{array}{ccc}
    0 & E_8 & E_9 \\
    E_8 & 0 & 0 \\
    E_9 & 0 & 0 \\
  \end{array} \right) \ ,   &
F_2& := F_2^\mu{}_\nu = \left(  \begin{array}{ccc}
    0 & \tilde E_8 & \tilde E_9 \\
    \tilde E_8 & 0 & 0 \\
    \tilde E_9 & 0 & 0 \\
  \end{array} \right) \ . \label{F12}
\end{align}
In general, there are two different cases to consider depending on whether the above field strengths $F_1$ and $F_2$ commute or not.
\begin{align}
[F_1,F_2]  &= 0  \quad : &&\text{corresponds to {\it parallel} electric fields}\footnotemark \ , \\
[F_1, F_2] & \neq 0 \quad : &&\text{corresponds to {\it oblique} electric fields}  \ . 
\end{align}\footnotetext{We include here also the case when one of the electric fields is zero.}
From a kinematic relativistic point of view there is a difference between the two cases. Indeed, suppose that one makes a boost of the system in a direction parallel to the electric field $E$;  then in the first case the system is invariant! However, in the second case, one sees effectively a magnetic field proportional to the cross product of the original electric fields $E \wedge \tilde E$. The second case would  thus correspond to the analogue of Thomas precession for strings. We shall focus first on the case of parallel electric fields in compact spaces. Then we discuss the non-parallel case.

Let us turn back to the boundary conditions \eqref{bc1}-\eqref{bc2}. A general solution to the wave equation is of the form
\be
X^\mu(\tau , \sigma) = X_L^\mu(\sigma_+) + X_R^\mu(\sigma_-) \ , 
\ee
where $\sigma_\pm = \tau \pm \sigma$. Then one can rewrite the boundary conditions in the convenient form
\begin{align}
({\bf 1} - F_1)^\mu{}_\nu\, \partial_+ X_L^\nu &= ({\bf 1} + F_1)^\mu{}_\nu\, \partial_- X_R^\nu \ ,  & \sigma &= 0  \ , \\
({\bf 1} + F_2)^\mu{}_\nu\, \partial_+ X_L^\nu &= ({\bf 1} - F_2)^\mu{}_\nu\, \partial_- X_R^\nu \ ,  & \sigma &= \pi \ . 
\end{align}
It is now natural to define the boost matrices $\Lambda_1$ and $\Lambda_2$ as the Cayley transforms of the field strengths
\begin{align}
\Lambda_i &= ({\bf 1} + F_i)^{-1} ({\bf 1} - F_i)  \ , & i&=1,2 \ .
\end{align}
Notice that the matrices above are well defined as long as $\det ({\bf 1} \pm F_i) \neq 0$. Already at this point one can see that there exists a critical electric field
\begin{equation}
||E_{cr}|| = 1 \quad \implies \quad \det ({\bf 1} \pm F_{cr}) = 1-||E_{cr}||^2 = 0 \ . 
\end{equation}
In toroidal compactification, models with electric fields correspond by T-duality to branes (of lower dimensionality) moving with velocities equal to the original electric fields. In the moving brane interpretation the critical value of the velocity corresponds to a motion at the speed of light $c=1$. Hence, the following condition will be satisfied by the electric fields
\begin{equation}
||E||<1 \quad \text{and} \quad ||\tilde E|| <1 \ . 
\end{equation}
Making use of the matrices above one can infer a mathematically equivalent form of the boundary conditions involving a periodicity condition for the left-moving coordinates $X_L^\mu$ together with a `twisted' identification of the left-moving with the right-moving
\begin{align}
\partial X_L^\mu(\tilde \tau + 2 \pi) &= (\Lambda_2 \Lambda_1)^\mu{}_\nu\, \partial X_L^\nu (\tilde \tau)  \ , \label{periodic}\\
\partial X_R^\mu(\tilde \tau) & = \Lambda_1^\mu{}_\nu \, \partial X_L^\nu(\tilde \tau) \label{L-R} \ , 
\end{align}
where $\tilde \tau \in \mathbb{R}$ is an arbitrary real variable. The periodicity property \eqref{periodic} is important as it determines the shifts of the frequencies in the mode expansions. They are determined by the eigenvalues of the matrix $\Lambda_2 \Lambda_1$. If $\lambda_a$ is an eigenvalue of $\Lambda_2 \Lambda_1$, the corresponding (imaginary) electric shift $i \epsilon_a$ is related to it by
\be
\lambda_a = e^{2 \pi \epsilon_a} \ . 
\ee
Furthermore, from the left-right identification in eq. \eqref{L-R} we see that the phase shifts in the mode expansions are determined by the matrix $\Lambda_1$.  When solving the boundary conditions there are two sub-cases to consider:
\begin{align}
F_1+ F_2& = 0 \quad : &&\text{corresponds to {\it dipole} strings} \ , \label{neutral}\\
F_1+F_2 & \neq 0 \quad : && \text{corresponds to {\it charged} strings} \ . 
\end{align}
The charged strings and dipole strings have different mode expansions, and we also sketch their canonical quantisation.  

We analyse in the next Section the case in which the electric fields $E$ and $\tilde E$ are parallel.  The case of {\it oblique} electric field is treated in Section \ref{s7}.

\section{Toroidal Compactification with Parallel Electric Fields}
\label{s6}
\subsection{Dipole Strings}

Dipole strings correspond to a `degenerate' case where the total charge of the open strings is equal to zero \eqref{dipole} and the boundary gauge potentials are identical. The first consequence of imposing \eqref{neutral} is that the frequency shifts in the mode expansions are vanishing. Indeed, it is easy to see that we have the identity
\begin{align}
\Lambda_2 = ({\bf 1} +F_2)^{-1}({\bf 1} - F_2) = ({\bf 1} - F_1)^{-1}({\bf 1} + F_1) = \Lambda_1^{-1} \ , 
\end{align}
hence the matrix whose eigenvalues determine the frequency shifts is the identity matrix
\begin{align}
\Lambda_2 \Lambda_1 = \Lambda_1^{-1} \Lambda_1 = {\bf 1} \ . \label{identity}
\end{align}
Thus, for dipole strings, the oscillator part of the mode expansion is identical to the standard one up to a phase determined by the eigenvalues of $\Lambda_1$, the matrix determining the left-right identification for open string in electric fields.
Indeed, making use of eq. \eqref{identity} into the boundary conditions in eqs. \eqref{periodic}-\eqref{L-R}, one can easily see that we must have
\begin{align}
\partial X_L^\mu (\sigma_+) &=\frac12 \sum_{n \in \mathbb{Z}} \alpha_n^\mu e^{-in\sigma_+}  \ ,  & \partial X_R^\mu (\sigma_-) &=\frac12 \sum_{n \in \mathbb{Z}} \Lambda_1^\mu{}_\nu \, \alpha_n^\nu e^{-in\sigma_-}  \ . 
\end{align}
Hence, after integration, we find the following general solution for the mode expansion of dipole strings:
\begin{align}
X^\mu(\tau, \sigma) &= x^\mu +\frac12 \alpha_0^\mu\, (\tau + \sigma) +\frac12 \Lambda_1^\mu{}_\nu\, \alpha_0^\nu\, (\tau - \sigma) + \frac{i}{2} \sum_{n \neq 0} \left[ \frac{\alpha_n^\mu}{n} e^{-in(\tau + \sigma)} + \Lambda_1^\mu{}_\nu \frac{\alpha^\nu_n}{n} e^{-in(\tau - \sigma)}\right] \ . 
\end{align}
Quantisation of dipole (and charged) strings  in magnetic fields has been first carried out in \cite{Abouelsaood:1986gd}. Our expression is similar to the ones found in \cite{Abouelsaood:1986gd} after analytic continuation. In order to see this, one needs to define coordinates $Y^a$ which diagonalise the boost matrix $\Lambda_1$. Due to the fact that we have chosen the electric field $E$ in an arbitrary direction in the plane $(X^8,X^9)$ the matrix that connects $Y^a$ to $X^\mu$ is factorised as a rotation $R_{\beta}$, which aligns the electric field $E$ with one of the axes, times a light cone change of variables $B:=B^\mu{}_a$. Hence we can define 
\begin{equation}
X^\mu = C^\mu{}_a\, Y^a \ , \quad \quad C^\mu{}_a = (R_\beta^{-1})^\mu{}_\nu\, B^\nu{}_a \ . \label{coordinates}
\end{equation}
Let us parametrise the (non-zero) components of the electric field as $E_8=||E||\cos \beta$ and $E_9 =  ||E|| \sin \beta$ such that $\beta$ is the angle between $E$ and the axis $X^8$. The matrices $R_\beta$ and $B$ are then given by
\begin{align}
R_\beta &=  \left( \begin{array}{ccc}
    1 & 0 & 0 \\
    0 & \cos \beta & \sin \beta \\
    0 & -\sin \beta & \cos \beta \\
  \end{array} \right) \ , & B = \frac{1}{\sqrt{2}} \left( \begin{array}{ccc}
    0 & -1 & 1 \\
    0 & 1 & 1 \\
    \sqrt{2} & 0 & 0 \\
  \end{array} \right) \ .  \label{RB}
\end{align}
With this choice, the initial Minkowski metric $\eta_{\mu \nu} = \text{diag}(-1,1,1)$ becomes a light cone one that we normalise as follows
\begin{align}
\eta_{ab} = \eta_{\mu \nu} C^\mu{}_a C^\nu{}_b = \left(   \begin{array}{ccc}
    1 & 0 & 0 \\
    0 & 0 & 1 \\
    0 & 1 & 0 \\
  \end{array} \right) \ . 
\end{align}
It is natural to label the directions $Y^a$ by the eigenvalues of the matrix $\Lambda_1$. Let us introduce the `rapidity' $\theta$ as being the norm of the electric field
\be
\tanh \theta :=||E|| \ . \label{rapid1}
\ee
In terms of $\theta$, we can write the eigenvalues of $\Lambda _1$ in the following way
\begin{align}
\Lambda_1^a{}_b:=(C^{-1})^a{}_\mu \, \Lambda_1^\mu{}_\nu \, C^\nu{}_b = \left( \begin{array}{ccc}
    1 & 0 & 0 \\
    0 & \frac{1+||E||}{1-||E||} & 0 \\
    0 & 0 & \frac{1-||E||}{1+||E||} \\
  \end{array} \right)
=  \left( \begin{array}{ccc}
    1 & 0 & 0 \\
    0 & e^{2 \theta} & 0 \\
    0 & 0 & e^{-2 \theta} \\
  \end{array} \right) \ ,
\end{align}
where the order of the eigenvalues is such that $a=(0,+,-)$. The coordinate $Y^0$ has a standard mode expansion corresponding to the direction orthogonal to the electric field in the plane $(X^8,X^9)$. On the other hand, the coordinates $Y^\pm$ have the form of the light cone coordinates found in \cite{Abouelsaood:1986gd}. Indeed, it results
\begin{align}
Y^0 & = y^0 + \alpha^0_0 \, \tau + i \sum_{n \neq 0} \frac{\alpha_n^0}{n} e^{-in\tau} \cos n\sigma \ ,  \label{ortho}\\
Y^\pm & = y^\pm + \frac{\tau \mp ||E|| \sigma}{1\mp ||E||} \alpha_0^\pm + i e^{\pm \theta} \sum_{n\neq 0} \frac{\alpha_n^\pm}{n} e^{-in \tau} \cos (n \sigma \mp i \theta) 
\ . \label{pm}
\end{align}
It is convenient to normalise the zero modes such that they have canonical Poisson brackets and commutators. From imposing the usual algebra for $X^\mu$ and its canonical conjugate momentum $P^\mu$ derived from eq. \eqref{action} (and containing the boundary terms) 
\begin{align}
\{X^\mu(\tau,\sigma), P^\nu(\tau, \sigma') \} &=\pi \eta^{\mu \nu} \delta(\sigma - \sigma') \ , \\
\{ X^\mu(\tau,\sigma), X^\nu(\tau,\sigma') \} & = \{ P^\mu(\tau,\sigma), P^\nu(\tau,\sigma') \} = 0 \ , 
\end{align}
one can infer that, in order to bring the zero mode algebra into its canonical form, the following rescaling/redefinition of the momenta $\alpha^\pm_0$ is necessary
\begin{align}
\frac{\alpha^\pm_0}{1\mp ||E||} \mapsto \frac{p^\pm}{1-||E||^2} \ . 
\end{align}
After introducing the notation $p^0:= \alpha_0^0$, we can thus write the usual commutation relations for the zero modes $y^a$ and $p^a$,
\begin{equation}
\{y^a , p^b\}  = \eta^{ab} \quad\quad \text{or} \quad\quad [y^a,p^b] = i \eta^{ab} \ . 
\end{equation}
Notice that the oscillators $\alpha_n^a$ satisfy the standard algebra for open strings. Since the quantisation is carried out in a covariant way, one needs to consider also the ghost fields associated to the gauge fixing of the local symmetries of the action in eq. \eqref{action}.
Some comments are in order about the choice of not aligning the electric field $E$ with any of the axes. In the case of non-compact directions $(X^8,X^9)$ this has no physical consequence. Indeed, one can always define a rotation which leaves the physical system invariant in such a way that (in the new coordinates) the electric field is aligned with one of the axes. However,
when compactifying on a two torus $\mathbb{T}^2$, only rotations of quantised angle are allowed! Let us first derive  the annulus in the case of a non-compact space. We can make use of the rotated coordinates
\be
\tilde X^\mu = (R_\beta)^\mu{}_\nu \, X^\nu \ , \label{rotcoord}
\ee
which have the property that the electric field $E$ is parallel to the axis $\tilde X^8$. Their mode expansions are then found to be
\begin{align}
\tilde X^0 &= X^0 = \tilde x^0+ \frac{1}{1-||E||^2}\left( \tilde q^0 \tau+ ||E||\, \tilde q^8 \sigma \right)+ \text{oscillators} \ , \label{rot0}\\
\tilde X^8 & =\tilde x^8+ \frac{1}{1-||E||^2}  \left(\tilde q^8\tau +||E||\, \tilde q^0 \sigma\right)  + \text{oscillators} \ , \label{rot8}\\
\tilde X^9 & = \tilde x^9 + \tilde q^9 \tau + \text{oscillators}  \ ,  \label{rot9}
\end{align}
with the zero modes $\tilde x^\mu$, $\tilde q^\mu$ satisfying the canonical commutation algebra
\be
[\tilde x^\mu, \tilde q^\nu] = i\eta^{\mu\nu} \ . 
\ee
In order to write the contribution of the zero modes to the annulus we choose a normalisation such that $L_0$ depends explicitly on the electric field. Indeed, the relevant part of $L_0$ is the following
\begin{align}
L_0^{zero} =\frac12\left[-\frac{1}{1-||E||^2}(\tilde q^0)^2 + \frac{1}{1-||E||^2} (\tilde q^8)^2 + (\tilde q^9)^2\right]  \label{L0dipole}
\end{align}
that gives rise to a contribution of the form
\begin{align}
\int d^3 \tilde x \, d^3 \tilde q \, e^{- \pi \tau_2 \, L_0^{zero}} =V_3\, (1-||E||^2) \frac{1}{\tau_2^{3/2}} \ , 
\end{align}
where the factor $1-||E||^2$ arises from the integration over the momentum zero modes $\tilde q^0$ and $\tilde q^8$. Hence, in the non-compact case, the annulus amplitude for the dipole strings is the standard one multiplied by the aforementioned factor. We recover in this way\footnote{Our partition functions in electric fields should be multiplied by $1/(2 \pi^2)^{D/2}$, where $D$ is the number of non-compact dimensions. We don't write explicitly this factor.}
 the result in \cite{Abouelsaood:1986gd}, 
\begin{align}
\mathcal{A}_d =\frac{V_{10}}2 (1-||E||^2)\int_0^\infty \frac{d\tau_2}{\tau_2^{6}}\sum_{\alpha,\beta}  c_{\alpha \beta}  \frac{\vartheta^4\left[\alpha \atop \beta \right] }{\eta^{12} } 
\ . 
\end{align}
In the following, we assume for simplicity a rectangular torus $\mathbb{T}^2$ such that the periodicities of the coordinates $X^{8,9}$ imply the usual quantisation of the (canonical) momentum modes $q^{8,9}$
\begin{align}
q^ 8 &= \frac{ m_8}{R_8} \ , & q^9 & = \frac{ m_9}{ R_9} \ . 
\end{align}
Since, in general, one cannot make use of the coordinates $\tilde X^\mu$ globally we pass back to the original $X^\mu$ which are not aligned with the electric field. Their mode expansions can be inferred from eqs. \eqref{rot0}-\eqref{rot9} to be
\begin{align}
X^0 &= x^0 + \frac{1}{1-||E||^2} \left[q^0 \tau + (E_8\, q^8 + E_9\, q^9) \sigma \right]+ \dots \quad , \\
X^8 &= x^8 +\frac{E_8(E_8\, q^8 + E_9\, q^9)}{||E||^2(1-||E|^2|)}\tau+ \frac{E_9(E_8\, q^9 - E_9\, q^8)}{||E||^2}\tau+\frac{E_8}{1-||E||^2}q^0 \sigma +\dots \quad , \\
X^9 &= x^9- \frac{E_9 (E_8\, q^8 + E_9\, q^9)}{||E||^2(1-||E|^2|)}\tau+ \frac{E_9 (E_8\, q^9 - E_9\, q^8)}{||E||^2}\tau- \frac{E_9}{1-||E||^2}q^0 \sigma +\dots \quad ,
\end{align}
where we have now defined the zero modes $x^\mu = (R_\beta^{-1})^\mu{}_\nu \, \tilde x^\nu$ and $q^\mu = (R_\beta^{-1})^\mu{}_\nu \, \tilde q^\nu$ such that their algebra remains canonical
\begin{align}
[x^\mu, q^\nu] = i \eta^{\mu \nu} \ . 
\end{align}
We can also write the relevant part of $L_0$ from eq. \eqref{L0dipole} in terms of the zero modes $x^\mu$ and $q^\mu$ as follows
\begin{align}
L_0^{zero} = \frac{1}{2(1-||E||^2)}\left[- (q^0)^2 + (1-E_9^2)\,(q^8)^2 + (1-E_8^2)\,(q^9)^2+2E_8\, E_9\, q^8\, q^9 \right] \ . 
\end{align}
Notice that the result above is consistent with the open string metric $G_{\mu \nu}$ defined in \cite{Seiberg:1999vs} since we have
\begin{align}
G_{\mu \nu}:= \eta_{\mu \nu} - (F_1 \eta^{-1} F_1)_{\mu \nu} = \left(  \begin{array}{ccc}
    -1+||E||^2 & 0 & 0 \\
    0 & 1-E_8^2 & -E_8\, E_9 \\
    0 & -E_8 \, E_9 & 1-E_9^2 \\
  \end{array} \right) \ . 
\end{align}
Indeed, the inverse of the metric above can be easily found
\begin{align}
G^{-1}_{\mu \nu} = \frac{1}{1-||E||^2}
 \left(  \begin{array}{ccc}
    -1 & 0 & 0 \\
    0 & 1-E_9^2 & E_8\, E_9 \\
    0 & E_8 \, E_9 & 1-E_8^2 \\
  \end{array} \right) \ , 
 \label{invmet}
 \end{align}
 such that $L_0^{zero}$ can be written the following form
 \begin{align}
 L_0^{zero} =\frac12 G^{-1}_{\mu \nu}\,  q^\mu \, q^\nu \ . 
 \end{align}
Turning back to the annulus, the momentum integration over $q^8$ and $q^9$ is replaced in the compact case by a sum over $m_8$ and $m_9$ in the usual way 
\be
\int dq^8 dq^9 \rightarrow \frac{1}{R_8R_9} \sum_{m_8,m_9 \in \mathbb{Z}} \ . \label{statistical}
\ee
Thus, from the zero modes one has the following contribution to the amplitude
\begin{align}
\int d^3 x \, d q^0 \, \frac{1}{R_8 R_9}\sum_{m_8, m_9} q^{L_0^{zero}} = V_1 \sqrt{1-||E||^2} \sum_{m_8,m_9 } e^{-  \frac{\pi \tau_2}{2(1-||E||^2)}\left[(1-E_9^2)\frac{m_8^2}{R_8^2} + (1-E_8^2) \frac{m_9^2}{R_9^2} + 2 E_8 E_9 \frac{m_8}{R_8}\frac{m_9}{R_9}\right] } .\label{sum1}
\end{align}
Finally, the annulus for dipole strings in the case of an electric field $E$ pointing into a compact two torus $\mathbb{T}^2$ is given by
\begin{align}
\mathcal{A}_d =\frac{V_{10}}2 \sqrt{1-||E||^2} \int_0^\infty \frac{d\tau_2}{\tau_2^{5}}\sum_{\alpha,\beta}  c_{\alpha \beta}  \frac{\vartheta^4\left[\alpha \atop \beta \right] }{\eta^{12}}\, P_{m_8,  m_9} \ , 
\end{align}
where $P_{m_8, m_9}$ denotes the sum over the momenta appearing in eq. \eqref{sum1}, with the convention to include the normalisation with the torus volume as in eq. \eqref{statistical}.  Let us now calculate the transverse channel amplitude with modular parameter $l =2/\tau_2$. Making use of the modular properties of the Jacobi $\vartheta$ -functions and of the Dedekind $\eta$ -function we can write the following amplitude for dipole strings in the tree-level (transverse) channel
\begin{align}
 \mathcal{\tilde A}_{d} &=2^{-5} \frac{V_{10}}{2}(1-||E||^2) \int dl\, \sum_{\alpha \beta} c_{\alpha \beta}  \frac{\vartheta^4\left[\alpha \atop \beta \right](il) }{\eta^{12}(il)}\, \tilde P_{ n_8,  n_9}  \ , \label{tr1}
\end{align}
where 
\begin{align}
\tilde P_{ n_8, n_9} = \sum_{ n_8,  n_9} e^{-\pi l \left[(1-E_8^2) n_8^2 R_8^2+(1-E_9^2)n_9^2 R_9^2 - 2 E_8 E_9\, n_8 n_9\, R_8 R_9 \right]}  
\end{align}
results from applying the Poisson summation formula to $P_{ m_8, m_9}$ and contains the $(8,9)$-bloc of the open string metric $G_{\mu \nu}$ of eq. \eqref{invmet}. One can see from eq. \eqref{tr1} that the factor $(1-||E||^2) $ is indeed consistent with the DBI effective action.
\subsection{Charged Strings}

Let us now derive  the mode expansions for charged open strings with parallel boundary electric fields. In this case the coordinates $Y^a$ defined in eq. \eqref{coordinates} are again a natural choice since they diagonalise, besides $\Lambda_1$, also the matrix $\Lambda_2 \Lambda_1$ appearing in the boundary condition \eqref{periodic}. This is due to the fact that $\Lambda_1$ and $\Lambda_2$ commute. Notice that this is no longer true in the case of oblique electric fields. Let us also define the `rapidity' $\tilde \theta$ related to the norm of $\tilde E$ by
\be
\tanh \tilde \theta :=||\tilde E||  \ . \label{rapid2}
\ee
The matrix product $\Lambda_2 \Lambda_1$ in the basis defined by coordinates $Y^a$ has the following diagonal form
\begin{align}
(\Lambda_2 \Lambda_1)^a{}_b = (C^{-1})^a{}_\mu\,  (\Lambda_2 \Lambda_1)^\mu{}_\nu\, C^\nu{}_b =   \left( \begin{array}{ccc}
    1 & 0 & 0 \\
    0 & e^{2(\theta + \tilde \theta)} & 0 \\
    0 & 0 & e^{-2(\theta + \tilde \theta)} \\
  \end{array} \right) \ . 
\end{align}
Thus, the electric field frequency shift $\epsilon$ is the sum of the `rapidities' $\theta$ and $\tilde \theta$ divided by a factor of $\pi$
\be
\epsilon:= \frac{1}{\pi}(\theta + \tilde \theta) \ , \label{epspar}
\ee
such that, from eq. \eqref{periodic}, the periodicities of $\partial Y_L^a$ result
\begin{align}
\partial Y_L^0(\tilde \tau + 2 \pi) &=\partial Y_L^0(\tilde \tau) \ , \label{ya1} \\
\partial Y_L^\pm (\tilde \tau + 2 \pi) &= e^{\pm2 \pi \epsilon} \partial Y_L^\pm (\tilde \tau) \ . \label{ya2}
\end{align}
Making use of the equations above together with eq. \eqref{L-R}, we can expand the derivatives of the left- and right-moving coordinates into modes
\begin{align}
\partial Y_L^\pm (\sigma_+) &=\frac12 \sum_{n \in \mathbb{Z}}  \alpha_{n\pm i \epsilon}^\pm\, e^{-i(n\pm i \epsilon) \sigma_+} \quad, & \partial Y_R^\pm (\sigma_-) & =\frac12 e^{\pm 2 \theta} \sum_{n \in \mathbb{Z}}  \alpha_{n\pm i \epsilon}^\pm\, e^{-i(n\pm i \epsilon) \sigma_-} \ . 
\end{align}
Finally, after integration, we obtain the known mode expansion \cite{Bachas:1992bh} for charged strings in parallel boundary electric fields
\begin{align}
Y^\pm(\tau ,\sigma) &=y^\pm + i\, e^{\pm \theta} \sum_{n \in \mathbb{Z}} \frac{\alpha_{n\pm i\epsilon}^\pm}{n\pm i \epsilon} e^{-in \tau} \cos [(n\pm i \epsilon) \sigma \mp i \theta]
\ , 
\end{align}
with $Y^0$ having the standard mode expansion already given in eq. \eqref{ortho}. Canonical quantisation leads to the following commutation algebra for the modes
\begin{align}
[y^+, y^-] &= \frac{i \pi}{||E+ \tilde E||} \ , & \left[\alpha^+_{n+i\epsilon}, \alpha^-_{m- i \epsilon} \right] = (n+i\epsilon) \, \delta_{n+m,0} \ , \label{paralg}
\end{align}
together with the standard commutation for the direction $Y^0$ defined to be orthogonal to the electric field $E$
\begin{align}
[y^0,p^0]=i \ . 
\end{align}
In terms of the rotated coordinates $\tilde X^\mu$ defined in eq. \eqref{rotcoord}, the mode expansions take the following simple form
\begin{align}
\tilde X^0 &= \tilde x^0 + \text{oscillators} \ , \\
\tilde X^8 & = \tilde x^8 + \text{oscillators} \ , \\
\tilde X^9 & = \tilde x^9 + \tilde p^9\, \tau + \text{oscillators} \ , 
\end{align}
where the zero mode $p^0\equiv \tilde p^9$ is now identified with the momentum along the direction $\tilde X^9$, thus orthogonal to  the parallel electric fields. The contribution to the annulus from the $\tilde x^\mu$ and $\tilde p^9$ can now be found to be
\begin{align}
\int \frac{d^3 \tilde x \, d\tilde p^9}{\sqrt{\det{\tilde \Omega}}}\,  q^{L_0^{zero}} = \frac{V_3}{\pi}\, ||E+ \tilde E|| \frac{1}{\tau_2^{1/2}} \ ,  
\end{align}
with the relevant part of $L_0$ given by $L_0^{zero} =(\tilde p^9)^2$ and the matrix $\tilde \Omega$ with determinant equal to $\pi^2/||E+ \tilde E||^2$ defined to encode the algebra of zero modes in the following way
\begin{equation}
\tilde x^A :=  \left( \begin{array}{cccc}
    \tilde x^0 &  \tilde x^8  &    \tilde x^9  &   \tilde p^9\\
  \end{array} \right)^T \ , \quad\quad [\tilde x^A,\tilde x^B]= \tilde \Omega^{AB} \ . 
\end{equation}
The inclusion of the factor with the square root of the determinant is necessary for a correct definition of the quantum volume. Finally, combining the contribution from the zero modes with the oscillators and the fermions, we can write the annulus in the non-compact case as
\begin{align}
\mathcal{A}_c =\frac{V_{10}}2\, ||E+ \tilde E|| \int_0^\infty \frac{d\tau_2}{\tau_2^{5}}\sum_{\alpha,\beta}  c_{\alpha \beta} \frac{\vartheta \left[ \alpha \atop \beta\right] \left( i\epsilon \frac{i\tau_2}{2}\Big |  \frac{i \tau_2}{2}\right)}{\vartheta \left[\alpha \atop \beta\right] \left ( 0 \Big | \frac{i\tau_2}{2}\right)} \frac{\vartheta^4\left[\alpha \atop \beta \right] \left(0 \Big | \frac{i\tau_2}{2}\right)}{\eta^{12} \left(\frac{i\tau_2}{2}\right)}\frac{i \eta^3 \left(\frac{i\tau_2}{2}\right)}{\vartheta_1\left ( i\epsilon \tfrac{i\tau_2}{2}|\tfrac{i\tau_2}{2}\right)} \ .  \label{chargednc}
\end{align}
Let us now turn to the case of a two torus $\mathbb{T}^2$ spanned by the coordinates $X^8$ and $X^9$. The use of the rotated $\tilde X^\mu$ is natural since the electric field $E$ is aligned with the axis $\tilde X^8$ and also since the charged string admits a momentum zero modes in the direction $\tilde X^9$. In principle one has to consider two cases depending on whether the direction $\tilde X^9$ is compact or not. In the first case one obtains a quantisation for the zero mode $\tilde p^9$ and subsequently a lattice sum in the annulus. In the second case one only has a standard momentum integration and the result is identical to the non-compact case in eq. \eqref{chargednc}. Let us investigate the condition for the coordinate $\tilde X^9$ to be compact. As before, we consider a rectangular two torus $\mathbb{T}^2$ with periodicities
\begin{equation}
X^8 \rightarrow X^ 8 + 2 \pi R_8 \ , \qquad\qquad X^9  \rightarrow X^9 + 2 \pi R_9  \ , \label{shift}
\end{equation}
such that the corresponding lattice is generated by the following orthogonal vectors
\begin{align}
\vec e_1 &= \left(  \begin{array}{c}
    2 \pi R_8 \\ 
    0 \\ 
  \end{array} \right)\ , &&  & \vec e_2 & = \left(  \begin{array}{c}
    0 \\ 
    2 \pi R_9 \\ 
  \end{array} \right)  \  . 
\end{align}
If the direction defined by the coordinate $\tilde X^9$ is compact,  there exists a lattice vector $\vec e_\perp$ that is parallel to it. This implies the existence of (coprime) integers $p,q$ such that
\be
\vec e_\perp = p\, \vec e_1 + q\, \vec e_2 \ . 
\ee
The condition for the integers to be coprime ensures that we choose a vector $e_\perp$ of minimal length. As a consequence of the above, the periodicity in the direction $\tilde X^9$ is determined by the length of the vector $| \vec e_\perp| := 2 \pi R_\perp$. If one defines the angle between $\vec e_\perp$ and $\vec e_1$ to be equal to $\beta + \pi/2$, then the quantisation condition for the angle $\beta$ is
\begin{align}
 \tan \beta & = -\frac{p R_8}{q R_9} =- \cot \left(\beta + \frac{\pi}{2}\right)\ ,  &&  & R_\perp^2 & = p^2 R_8^2 + q^2 R_9^2  \ .  \label{qangle}
\end{align} 
Notice that an additional condition is necessary in order for $\tilde X^8$ to be compact as well. Indeed, in this case there exists a lattice vector $\vec e_\parallel$ parallel to $\tilde X^8$ defined by coprime integers $l,k$
\be
\vec e_\parallel = k \, \vec e_1 + l \, \vec e_2 \ . 
\ee
As before, the length of the vector $|\vec e_\parallel| :=2 \pi R_\parallel$ determines the periodicity in the direction $\tilde X^8$ whereas the (same) angle $\beta$ satisfies a different quantisation condition
\begin{align}
R_\parallel^2 & = k^2 R_8^2 + l^2 R_9^2 \ , &&   & \tan \beta & = \frac{l R_9}{k R_8} \  .   \label{qanglep}
\end{align} 
If one combines the two quantisation conditions for the angle $\beta$ in eqs. \eqref{qangle}, \eqref{qanglep},a constraint on the radii
\be
\frac{R_8^2}{R_9^2} = -\frac{q\, l}{p\, k} \in \mathbb{Q} 
\ee 
comes out. It ensures that the two vectors $\vec e_\parallel$ and $\vec e_\perp$ are indeed orthogonal $\vec e_\parallel \cdot \vec e_\perp = 0$.

If we assume that $\tilde X^9$ is compact, then the usual quantisation condition\footnote{This is due to the fact that one has a plane wave solution in the $\tilde X^9$ direction.} for the momentum mode $\tilde p^9$ holds
\begin{align}
\tilde X^9 &\rightarrow \tilde X^9 + 2 \pi R_{\perp} \ , &&   & \tilde p^9 &= \frac{\tilde m_9}{R_{\perp} } \ . 
\end{align}
Hence, the momentum integration is replaced by the standard lattice sum 
$$P_{\tilde m_9} = \frac{1}{R_{\perp}} \sum_{\tilde m_9} e^{-\pi \tau_2 \tilde m_9^2/2R_{\perp}^2}$$ 
for electric fields such that $\tilde X^9$ is compact. Finally, in the compact $\tilde X^9$ case the annulus can be written as
\begin{align}
\mathcal{A}_c =\frac{V_{10}}2\, ||E+ \tilde E|| \int_0^\infty \frac{d\tau_2}{\tau_2^{9/2}}\sum_{\alpha,\beta}  c_{\alpha \beta} \frac{\vartheta \left[ \alpha \atop \beta\right] \left( i\epsilon \frac{i\tau_2}{2}\Big |  \frac{i \tau_2}{2}\right)}{\vartheta \left[\alpha \atop \beta\right] \left ( 0 \Big | \frac{i\tau_2}{2}\right)} \frac{\vartheta^4\left[\alpha \atop \beta \right] \left(0 \Big | \frac{i\tau_2}{2}\right)}{\eta^{12} \left(\frac{i\tau_2}{2}\right)}\frac{i \eta^3\left(\frac{i\tau_2}{2}\right)}{\vartheta_1\left ( i\epsilon \tfrac{i\tau_2}{2}|\tfrac{i\tau_2}{2}\right)}\, P_{\tilde m_9}\ .
\end{align}
In the original coordinates $X^\mu$ the momentum $\tilde p^9$ has non-zero projections on both the $X^8$ and $X^9$ axes that we denote by $p^8$ and $p^9$. However, the electric field has lifted the zero mode in the direction $\tilde X^8$, hence we can write the following relation
\begin{align}
\left(  \begin{array}{c}
    0 \\
    \frac{\tilde m_9}{\tilde R_9} \\
  \end{array} \right) = \frac{1}{||E+ \tilde E||}  \left(  \begin{array}{cc}
    E_8+\tilde E_8 & E_9 + \tilde E_9 \\
    -(E_9 + \tilde E_9) & E_8 + \tilde E_8 \\
  \end{array} \right) \left(  \begin{array}{c}
    \frac{m_8}{R_8} \\
    \frac{m_9}{ R_9} \\
  \end{array} \right) \ , 
\end{align}
where the matrix above is equal to $R_\beta$ in the case of parallel electric fields. The presence of the electric field imposes the constraint $\tilde p^8 = 0$. Then one can see the angle quantisation as a selection rule reducing the lattice sum to be only over $\tilde m_9$
\begin{align}
\frac{E_9}{E_8} =\frac{\tilde E_9}{\tilde E_8}= \tan \beta =- \frac{m_8 R_9}{m_9R_8} \ .  \label{angleq}
\end{align}

\subsection{Open/Closed String Duality}

A consistency check of the charged string annulus amplitudes derived in the case of parallel electric fields can be done by looking at the transverse channel interpretation as closed strings exchanged between branes. Indeed, making the change of variables
\begin{align}
l &= \frac{2}{\tau_2} \ ,  & \int_0^\infty dl = 2 \int_0^\infty \frac{d\tau_2}{\tau_2^2} \ , \label{transverse}
\end{align}
 one finds the following form of the transverse channel annulus for charged strings with compact $\tilde X^9$
\begin{align}
  \mathcal{\tilde A}_c &=2^{-5}\frac{V_{10}}2\, ||E+ \tilde E||\,  \int_0^\infty dl \sum_{\alpha,\beta}  c_{\alpha \beta} \frac{\vartheta \left[ \alpha \atop \beta\right] \left( i\epsilon  |  il \right)}{\vartheta \left[\alpha \atop \beta\right] \left ( 0 | il \right)} \frac{\vartheta^4\left[\alpha \atop \beta \right] \left(0  | il \right)}{\eta^{12} \left(i l\right)}\frac{i \eta^3(il)}{\vartheta_1\left ( i\epsilon |il\right)}\, \tilde P_{\tilde n_9} \ , 
\end{align}
with the dual lattice sum $\tilde P_{\tilde n_9}$ being
\begin{align}
\tilde P_{\tilde n_9} = \sum_{\tilde n_9} e^{- \pi l\, \tilde n_9^2  R_{\perp}^2} =  \sum_{\tilde n_9} e^{- \pi l\, \tilde n_9^2  (q^2 R_8^2 + p^2 R_9^2)}  \ . 
\end{align}
Notice that whereas this expression depends explicitly on the comprime integers $p,q$ defining the direction of the electric field, small changes in the angle are possible only if the integers are very large. In this case, the sum collapses to the first term ($\tilde n_9=0$), which turns out to be the result obtained in the non-compact case, where the torus volume is infinite. In this sense, the lattice sum and therefore the whole partition function is continuous under infinitesimal changes in the direction of the electric field \footnote{We thank Costas Bachas for pointing out to us the continuous behaviour of the partition function under small changes of the angle as a consistency check of the result.}. Taking the limit $p,q \rightarrow \infty$ with fixed ratio can also be understood as taking the limit of non-compact direction $\tilde X^9$: in this case the lattice sum is reduced to the standard result of continuous momentum integration. Notice, however, that open-closed string duality is consistent irrespective of the quantisation condition  (\ref{qanglep}) related to the compactness of the parallel coordinate, which is valid non-perturbatively by the arguments of Sections \ref{s2} and \ref{s3}.  Moreover, the partition function depends only on the compactness of the transverse coordinate, which is undetermined both perturbatively and non-perturbatively.
      
In order to show that we have the correct DBI interpretation, one needs to take into account the factor of $\sinh \pi \epsilon $ arising from the $\vartheta_1(i\epsilon|i l )$. Indeed, it is easy to obtain the identity
\begin{align}
\sinh \pi \epsilon = \frac{||E+\tilde E||}{\sqrt{(1-||E||^2)(1-||\tilde E||^2)}}  \  , 
\label{dbiiden} 
\end{align}
valid (only) for parallel electric fields. Using \eqref{dbiiden} we now have the following behaviour of the annulus in the transverse channel
\begin{align}
\mathcal{\tilde A}_c \sim \sqrt{(1-||E||^2)(1-||\tilde E||^2)} \ , 
\end{align}
consistent with the closed string interpretation.

\section{Oblique Electric Fields}
\label{s7}
\subsection{Mode Expansions}

Models with several stacks of branes will contain, in general, open strings with non-parallel boundary electric fields. The form of the boundary condition in eqs. \eqref{periodic}-\eqref{L-R} is particularly useful in this case. It is natural to work with coordinates that diagonalise the composition of boosts $\Lambda_2 \Lambda_1$ appearing in the boundary conditions. For this purpose we define $Y^a$ related to $X^\mu$ as follows
\be
X^\mu = C^\mu{}_a\, Y^a \ . 
\ee
Notice that the coordinates $Y^a$ are different than the ones used in the parallel case, but reduce to them in this limit. The matrix (of eigenvectors of $\Lambda_2\Lambda_1$) $C:=C^\mu{}_a$ is chosen to satisfy the identities
\begin{align}
(\Lambda_2 \Lambda_1)^\mu{}_\nu \, C^\nu{}_a & = \lambda_a \, C^\mu{}_a \ ,   & \eta_{ab} &= \eta_{\mu \nu} C^\mu{}_a C^\nu{}_b = \left(   \begin{array}{ccc}
    1 & 0 & 0 \\
    0 & 0 & 1 \\
    0 & 1 & 0 \\
  \end{array} \right) \ . 
\end{align}
Our procedure of solving the boundary conditions works only for diagonalisable matrices $\Lambda_2\Lambda_1$. However, this is always the case for sub-critical electric fields. We restrict as before, for simplicity, the electric fields $E$ and $\tilde E$ to lie on the plane $(X^8, X^9)$. It turns out that the eigenvalues of $\Lambda_2 \Lambda_1$ are given by
\begin{align}
(\Lambda_2 \Lambda_1)^a{}_b = (C^{-1})^a{}_\mu \, (\Lambda_2 \Lambda_1)^\mu{}_\nu \, C^\nu{}_b =  \left(   \begin{array}{ccc}
    \lambda_0 & 0 & 0 \\
    0 & \lambda_+ & 0 \\
    0 & 0 & \lambda_- \\
  \end{array} \right) = \left(   \begin{array}{ccc}
    1 & 0 & 0 \\
    0 & e^{2 \pi \epsilon} & 0 \\
    0 & 0 & e^{-2 \pi \epsilon} \\
  \end{array} \right) \ , 
\end{align}
where we have defined the electric shift $\epsilon:=\tfrac{1}{2\pi} \log \lambda_+$. The eigenvalue $\lambda_+$ has the following expression \footnote{The other eigenvalue different from $1$ is given by $\lambda_- = 1/\lambda_+$.}
\be
\lambda_+ = \frac{1}{(1-||E||^2)(1-||\tilde E||^2)} \left[(1+E^T \tilde E) + \sqrt{||E+ \tilde E||^2- ||E \wedge \tilde E ||^2} \right]^2 \ . 
\ee
Notice that the eigenvalues $\lambda_a$ are always real for sub-critical electric fields. Indeed, one can show that we have the implication\footnote{It is useful to make use of the following identity
 \begin{align}
||E+ \tilde E||^2 - ||E \wedge \tilde E||^2=(1+E^T \tilde E)^2-(1-||E||^2)(1-||\tilde E||^2) \ , 
\end{align}
\hspace{0.5 cm} valid for arbitrary vectors $E$ and $\tilde E$.}
\begin{equation}
||E||,||\tilde E||<1 \qquad \implies \qquad ||E+ \tilde E||^2- ||E \wedge \tilde E ||^2 \geq 0 \ . 
\end{equation}
Making use of the `rapidities' $\theta$ and $\tilde \theta$ defined in eqs. \eqref{rapid1} and \eqref{rapid2} and of the angle $\alpha = \sphericalangle(E,\tilde E)$ between the electric fields, we can write the following expression for the electric shift
\begin{align}
\pi \epsilon = \cosh^{-1}\left( \cosh \theta \cosh \tilde \theta +\cos \alpha \sinh \theta \sinh \tilde \theta\right) \ . \label{e-oblique}
\end{align}
The parallel case in eq. \eqref{epspar} is easily recoverable after setting $\cos \alpha = 1$.  It is interesting to find the Cayley generator, that we denote by $P$, of the product of boosts $\Lambda_2 \Lambda_1$. One can show that we have\footnote{Notice that the product in the reversed order is generated by the transposed of $P$, i.e. $$\Lambda_1 \Lambda_2 = ({\bf 1} + P^T)^{-1} ({\bf 1} - P^T) $$}
\begin{align}
\Lambda_2 \Lambda_1 & = ({\bf 1} +P)^{-1} ({\bf 1} -P)\ ,  & P &= \frac{1}{1+E^T \tilde E} \left(F_1 + F_2 + [F_1,F_2] \right) \ . 
\end{align}
From the form of $P$ one can infer, as expected,  the presence of the Thomas precession when the commutator $[F_1,F_2]$ is different from zero. The periodicity conditions for the coordinates $Y^a$ have the same form as the ones for parallel electric fields in eqs. \eqref{ya1}-\eqref{ya2}. However, in the oblique case, due to the fact that $\Lambda_1$ and $\Lambda_2$ do not commute, the matrix identifying the left-moving with the right-moving is no longer diagonal! We again split the index $a$ relative to the eigenvalues of $\Lambda_2 \Lambda_1$ such that we have
\be
a=(0,+,-) \ . 
\ee
Using  eqs. \eqref{ya1}-\eqref{ya2}, it is easy to show that the mode expansions for the derivatives $\partial Y^a_L$ are 
\begin{align}
\partial Y_L^0(\sigma_+) &= \frac12\sum_{n \in \mathbb{Z}} \alpha_n^0\, e^{-in \sigma_+} \ ,  & \partial Y_L^\pm(\sigma_+) &= \frac12\sum_{n\in \mathbb{Z}} \alpha^\pm_{n\pm i \epsilon}\, e^{-i (n\pm i\epsilon)\sigma_+}
\end{align}
and, using also eq. \eqref{L-R}, the right-moving part comes out to be
\begin{align}
\partial Y^a_R & =\frac12 \sum_{n \in \mathbb{Z}} \left[\Lambda_1^a{}_0 \, \alpha_n^0\, e^{-i n\sigma_-} + \Lambda_1^a{}_+ \, \alpha^+_{n+i\epsilon}\, e^{-i(n+i\epsilon) \sigma_-}  +\Lambda_1^a{}_- \, \alpha^-_{n-i\epsilon} \,e^{-i(n-i\epsilon) \sigma_-}\right] \ . 
\end{align}
Finally, after integration, the mode expansions for open strings with boundary oblique electric fields result
\begin{align}
   Y^a (\tau,\sigma)&=y^a + \frac12 \delta^a{}_0\, p^0(\tau+\sigma) +\frac12 \Lambda_1^a{}_0\, p^0 (\tau-\sigma) +\frac{i}{2}\sum_{n\neq 0} \frac{\alpha_n^0}{n} \left(\delta^a{}_0\, e^{-in(\tau+\sigma)} + \Lambda_1^a{}_0\, e^{-in(\tau-\sigma)} \right)  \nonumber\\
     & +\frac{i}2 \sum_{n\in \mathbb{Z}} \frac{\alpha^+_{n+i\epsilon}}{n+i\epsilon} \left[\delta^a{}_+\, e^{-i(n+i\epsilon)(\tau+\sigma)} + \Lambda_1^a{}_+\, e^{-i(n+i\epsilon)(\tau-\sigma)} \right] \nonumber\\
     &+\frac{i}2 \sum_{n\in \mathbb{Z}} \frac{\alpha^-_{n-i\epsilon}}{n-i\epsilon} \left[\delta^a{}_-\, e^{-i(n-i\epsilon)(\tau+\sigma)} + \Lambda_1^a{}_-\, e^{-i(n-i\epsilon)(\tau-\sigma)} \right]  \ . 
 \label{modexp}
\end{align}  
$L_0$ has a form similar to the case of parallel electric fields, 
\begin{align}
L_0 = \frac12 (p^0)^2 + \frac12 \sum_{n\neq 0} \alpha_{-n }^0 \alpha_n^0 + \sum_{n\in \mathbb{Z}} \alpha^+_{n+i \epsilon} \alpha^-_{-n-i\epsilon} +\frac12 i\epsilon(1-i\epsilon)
\ ,
\end{align}
but with frequency shift $\epsilon$ given in eq. \eqref{e-oblique}. 
Going back to the original coordinates $X^\mu$ is done with the matrix $C:=C^\mu{}_a$ that can be again factorised as a product of a rotation $R_\gamma$, which aligns the vector $E+ \tilde E$ with one of the axes, times an electric field dependent matrix $B:=B^\mu{}_a$ which transforms the usual Minkowski metric $\eta_{\mu \nu}$ into the light-cone metric $\eta_{ab}$.
Due to the fact that the Cayley transform of $\Lambda_2 \Lambda_1$ is proportional to $F_1 + F_2 + [F_1,F_2]$, one can use this latter  to find the matrix of eigenvectors $C$. After some algebra one finds 
\begin{align}
C^\mu{}_a = (R_\gamma^{-1})^\mu{}_\nu \, B^\nu{}_a \ , 
\end{align}
with the matrix $B$ given by
 \begin{align}
B^\mu{}_a &=\frac{1}{\sqrt{2(||E+\tilde E||^2 - ||E \wedge \tilde E||^2)}} \times \nonumber\\
&\left(   \begin{array}{ccc}
    -\sqrt{2}\,(E_8 \tilde E_9 - \tilde E_8 E_9) & -||E+\tilde E|| & ||E+\tilde E|| \\
    0 & \sqrt{||E+\tilde E||^2 - ||E \wedge \tilde E||^2} & \sqrt{||E+\tilde E||^2 - ||E \wedge \tilde E||^2} \\
    \sqrt{2}\,||E+\tilde E|| & E_8 \tilde E_9 - \tilde E_8 E_9 & -(E_8 \tilde E_9 - \tilde E_8 E_9) \\
  \end{array} \right) \label{bot}
\end{align}
and $R_\gamma$ having the same form as the rotation matrix in eq. \eqref{RB} but with $\gamma$ now being the angle between the vector $E+\tilde E$ and the axis $X^8$. The matrices $R_\gamma$ and $B$ (and hence also $C$) above reduce to eq. \eqref{RB} when the limit of parallel electric fields is taken, i.e. after setting $E \wedge \tilde E = 0$.

\subsection{Quantisation and the Annulus}

Standard canonical quantisation requires inverting the mode expansions \eqref{modexp} for the Fourier coefficients. Due to the fact that the matrix $\Lambda_1$ is not diagonal in the coordinates $Y^a$ one needs to study the orthogonality properties of the following matrix valued functions
\be
(f_n)^a{}_b(\tau,\sigma) =
\frac12\left\{
\begin{split}
&\delta^a{}_b\, e^{-i(n+i\epsilon_b)(\tau+\sigma)} + \Lambda_1^a{}_b\, e^{-i(n+i\epsilon_b)(\tau - \sigma)}\, , \qquad n+i\epsilon_b \neq 0\\
& \delta^a{}_0 (\tau+\sigma) + \Lambda_1^a{}_0 (\tau-\sigma)\, , \qquad n+i\epsilon_b = 0\\
\end{split}\right.
\ee
where, for convenience, we are using a covariant notation for the electric shifts $\{ \epsilon_a \} = \{\epsilon_0, \epsilon_+, \epsilon_- \} = \{0,\epsilon, - \epsilon \}$. In our notation the zero mode part corresponds to the function $(f_0)^a{}_0$. It is useful to introduce the following pairing between the functions $f_n$
\begin{align}
(f_m|f_n)_{ab}:=\int_0^\pi d\sigma\, \eta_{cd}\, (f_m)^c{}_a\, \overset\leftrightarrow{\partial}_\tau (f_n)^d{}_b + F_1{}_{cd}\, (f_m)^c{}_a\, (f_n)^d{}_b \Big|_{\sigma = 0} +  F_2{}_{cd}\, (f_m)^c{}_a\, (f_n)^d{}_b \Big|_{\sigma = \pi} \ , \label{pairing}
\end{align}
where we have defined the differential operator $(f_m)^c{}_a  \overset\leftrightarrow{\partial}_\tau (f_n)^d{}_b:= (f_m)^c{}_a \, \partial_\tau (f_n)^d{}_b - \partial_\tau (f_m)^c{}_a (f_n)^d{}_b  $.  In order to invert the mode expansions, one can use the following identities 
\begin{align}
(f_m|f_n)_{ab} &= i \pi (m+i\epsilon_a)\, \eta_{ab}\, \delta_{m+n,0}\, \delta_{\epsilon_a+\epsilon_b,0} \ , &(f_m|f_0)_{a0}  &= 0  \ , & (f_m|\delta)_{ab} & = 0 \ , \label{pair1}\\
(f_0| \delta)_{0b} &= - \pi \left(\eta - F_2\right)_{0b} \ , & (f_0|f_0)_{00} &= 0  \ , & (f_0| f_n)_{0b}  &=0  \ , \label{pair2} \\
(\delta| f_0)_{a0} &=\pi \left(\eta + F_2\right)_{a0} \ ,   & (\delta| \delta)_{ab}  &= (F_1+F_2)_{ab} \ , & (\delta| f_n)_{ab} & = 0 \ . \label{pair3}
\end{align}
Indeed, the mode expansions in eq. \eqref{modexp} can be written now with the help of the functions $f_n$ as 
\begin{align}
Y^a(\tau,\sigma) = y^a + (f_0)^a{}_0(\tau,\sigma) \, p^0 + i \sum_{n \in \mathbb{Z} \atop n+i\epsilon_b \neq 0} (f_n)^a{}_b(\tau,\sigma)\, \frac{\alpha^b_{n+ i\epsilon_b}}{n+i \epsilon_b} \ . 
\end{align}
It is now easy to see that the oscillators $\alpha^a_{n+i\epsilon_a}$ admit an integral representation in terms of $Y^a$ and its time derivative $\partial_\tau Y^a$ of the form
\begin{align}
\alpha^a_{n+i \epsilon_a} = \frac1{\pi} \eta^{ab} (f_{-n}|Y)_b =\frac1\pi \eta^{ab} &\left[\int_0^\pi d\sigma \, \eta_{cd}\, (f_{-n})^c{}_b \overset\leftrightarrow{\partial}_\tau Y^d +F_1{}_{cd}\, (f_{-n})^c{}_b\, Y^d \Big|_{\sigma = 0} \right. \nonumber \\ &+  \left. F_2{}_{cd}\, (f_{-n})^c{}_b\, Y^d \Big|_{\sigma = \pi} \right] \ .
\end{align}
The commutation algebra of the oscillators is similar to the one in the parallel case \eqref{paralg} but with the electric shift $\epsilon$ modified to the expression in \eqref{e-oblique}. Indeed, one can write the result in a covariant manner as follows
\begin{align}
\left[\alpha^a_{m+i\epsilon_a}, \alpha^b_{n+i\epsilon_b} \right] = \eta^{ab} (m+i\epsilon_a) \delta_{m+n,0}\, \delta_{\epsilon_a+ \epsilon_b,0} \ . 
\end{align}
The pairing that we defined in eq. \eqref{pairing} suggests the introduction of the following zero-mode redefinitions
\begin{align}
x_a &:= (\delta|Y)_a = \int_0^\pi d\sigma \, \eta_{cd}\, \delta^c{}_a \overset\leftrightarrow{\partial}_\tau Y^d +F_1{}_{cd}\, \delta^c{}_a\, Y^d \Big|_{\sigma = 0} +  F_2{}_{cd}\, \delta^c{}_a\, Y^d \Big|_{\sigma = \pi} \ , \label{rep1} \\
\pi_0&:=(f_0|Y)_0=\int_0^\pi d\sigma \, \eta_{cd}\, (f_0)^c{}_0 \overset\leftrightarrow{\partial}_\tau Y^d +F_1{}_{cd}\, (f_0)^c{}_0\, Y^d \Big|_{\sigma = 0} +  F_2{}_{cd}\, (f_0)^c{}_0\, Y^d \Big|_{\sigma = \pi} \ ,  \label{rep2}
\end{align}
which, after the use of the identities in eqs. \eqref{pair1}-\eqref{pair2}, can be shown to be related to the original modes $y^a$ and $p^0$ by the relations
\begin{align}
x_a &= (F_1+F_2)_{ab}\, y^b + \pi (\eta + F_2)_{a0}\, p^0 \ ,  \label{xa}\\
\pi_0 & = - \pi (\eta - F_2)_{0a}\, y^a \ . \label{pi0}
\end{align}
Making use of the integral representations in eqs. \eqref{rep1}-\eqref{rep2} one can compute the commutators of the modes $x_a$ and $\pi_0$. The result that one obtains has the simple form
\begin{align}
[x_a, x_b ] &=i \pi (F_1 + F_2)_{ab} \ , &  [ x_a, \pi_0 ] & =i \pi^2 (\eta+ F_2)_{a0}  \ . \label{xpialg}
\end{align}
With eqs. \eqref{xa}-\eqref{xpialg} at our disposal, we can now derive the commutation algebra for the modes $y^a$ and $p^0$.
The result is 
\begin{align}
[y^+,y^-] & = \frac{i \pi}{\sqrt{||E+\tilde E||^2 - ||E \wedge \tilde E||^2}} \ , && & [y^\pm,y^0] & = \pm \frac{i \pi F_2^{0\pm}}{\sqrt{||E+\tilde E||^2 - ||E \wedge \tilde E||^2}} \ , \\
[y^\pm,p^0]&= \mp \frac{i (F_1+F_2)^{0\pm}}{\sqrt{||E+\tilde E||^2 - ||E \wedge \tilde E||^2}} \ , &&   & [y^0,p^0] &= -\frac{i(F_1 +F_2)^{+-}}{\sqrt{||E+\tilde E||^2 - ||E 
\wedge \tilde E||^2}} \ . 
\end{align}
It is convenient to work with rotated coordinates $\tilde X^\mu$ defined such that the vector $E+ \tilde E$ is aligned with $\tilde X^8$
\begin{align}
\tilde X^\mu &= (R_\gamma)^\mu{}_\nu \, X^\nu \ , & \tilde X^\mu = B^\mu{}_a \, Y^a  \ . \label{tildeX}
\end{align}
The mode expansions of these coordinates have the following form
\begin{align}
\tilde X^0 &= \tilde x^0 - \frac{E_8 \tilde E_9 - \tilde E_8 E_9}{\sqrt{||E+ \tilde E||^2 - ||E \wedge \tilde E||^2}} \, p^0 \sigma + \text{oscillators} \ , \\
\tilde X^8 &= \tilde x^8 + \text{oscillators} \ , \\
\tilde X^9 & = \tilde x^9 + \frac{||E+ \tilde E||}{\sqrt{||E+ \tilde E||^2 - ||E \wedge \tilde E||^2}}\, p^0 \tau + \text{oscillators} \ ,
\end{align}
where the constant modes $\tilde x^\mu$ are related to the $y^a$ by the same matrix $B^\mu{}_a$ in eq. \eqref{bot}, hence
\be
\tilde x^\mu = B^\mu{}_a \, y^a \ . 
\ee
Since the vector $E+ \tilde E$ is aligned with $\tilde X^8$, the momentum zero mode $p^0$ is associated to the direction orthogonal to $E+ \tilde E$, namely $\tilde X^9$. In order to define properly the momentum integration one needs to find the correct normalisation from the commutator algebra of the zero modes. It turns out that the momentum $p^0$ has the following commutators
\begin{align}
[\tilde x^0, p^0] & = 0 \ , & [\tilde x^8, p^0] & = 0 \ , & [\tilde x^9, p^0] &= \frac{i||E+ \tilde E||}{\sqrt{||E+ \tilde E||^2 - ||E \wedge \tilde E||^2}} \ , 
\end{align}
whereas the $\tilde x^\mu$ satisfy a non-commutative spacetime algebra of the form
\begin{align}
[\tilde x^0, \tilde x^8] & = - \frac{i \pi}{||E+ \tilde E||} \ , & [\tilde x^0,\tilde x^9] &= -\frac{i \pi \tilde E^T(E+ \tilde E)(E_8 \tilde E_9 - \tilde E_8 E_9)}{||E+ \tilde E||\, (||E+ \tilde E||^2 - ||E \wedge \tilde E||^2)}  \ , \label{nc1}\\
 [\tilde x^8, \tilde x^9] &= 0 \ . \label{nc2}
\end{align}
We are thus led to the introduction of the canonically normalised momentum $\tilde p^9$, related to $p^0$ by the following expression:
\begin{align}
\tilde p^ 9 = \frac{\sqrt{|E+ \tilde E||^2 - ||E \wedge \tilde E||^2}}{||E+ \tilde E||} \, p^0 \ . 
\end{align}
We now have all the ingredients we need in order to define the annulus both in the compact and the non-compact case. The mode expansion for $\tilde X^9$ can be written as
\begin{align}
\tilde X^9 = \tilde x^9 + \frac{||E+ \tilde E||^2}{||E+ \tilde E||^2 - ||E \wedge \tilde E||^2} \, \tilde p^9\tau + \text{oscillators} \ ,
\end{align}
with the following algebra 
\begin{align}
[\tilde x^0, \tilde p^9] & = 0 \ , & [\tilde x^8, \tilde p^9] & = 0 \ , & [\tilde x^9, \tilde p^9] &= i \ , \label{canonical}
\end{align}
satisfied by $\tilde p^9$.  Since the zero modes $\tilde x^\mu$ and $\tilde p^9$ span a non-commutative algebra that is not in canonical form, one needs to divide the integration measure by the corresponding pfaffian (which will correctly define the quantum volume!). Indeed, let us introduce the notation
\begin{align}
\tilde x^A& =  \left( \begin{array}{cccc}
    \tilde x^0  & \tilde x^8 & \tilde x^9 & \tilde p^9 \\
  \end{array} \right)^T & \text{and} && [\tilde x^A, \tilde x^B]&=:\Omega^{AB} \ ,
  \end{align}
where the antisymmetric matrix $\Omega$ can be read off from eqs. \eqref{nc1}, \eqref{nc2} and \eqref{canonical}. Its determinant is given by $\det \Omega = \pi^2/||E+ \tilde E||^2$. Thus, we can further write that the correct measure in the annulus is given by
\begin{align}
\text{Measure} = \frac{d^3 \tilde x\, d\tilde p^9 }{\sqrt{\det \Omega}} = \frac{V_3}{\pi}||E+ \tilde E|| \, d\tilde p^9 \ . 
\end{align}
The integration over $\tilde x^\mu$ can be performed since $L_0$ only depends on $\tilde p^9$. In the case of a non-compact space-time $L_0$ has the form
\be
L_0^{zero} = \frac12(p^0)^2 = \frac{||E+ \tilde E||^2}{||E+ \tilde E||^2 - ||E \wedge \tilde E||^2} \, \frac12 (\tilde p^9)^2 \ . 
\ee
After putting together the other contributions from the oscillators and the orthogonal coordinates, one obtains the annulus amplitude for charged strings in the presence of oblique electric fields
\begin{align}
\mathcal{A} =\frac{V_{10}}2 \sqrt{||E+ \tilde E||^2 - ||E \wedge \tilde E||^2}\int_0^\infty \frac{d\tau_2}{\tau_2^{5}}\sum_{\alpha,\beta}  c_{\alpha \beta} \frac{\vartheta \left[ \alpha \atop \beta\right] \left( i\epsilon \frac{i\tau_2}{2}\Big |  \frac{i \tau_2}{2}\right)}{\vartheta \left[\alpha \atop \beta\right] \left ( 0 \Big | \frac{i\tau_2}{2}\right)} \frac{\vartheta^4\left[\alpha \atop \beta \right] \left(0 \Big | \frac{i\tau_2}{2}\right)}{\eta^{12} \left(\frac{i\tau_2}{2}\right)}\frac{i \eta^3}{\vartheta_1\left ( i\epsilon \tfrac{i\tau_2}{2}|\tfrac{i\tau_2}{2}\right)}. \label{annc}
\end{align}
Let us now compactify on a two torus $\mathbb{T}^2$ spanned by the coordinates $X^8, X^9$. Again, as in the parallel case, we have two cases to consider depending on whether the direction orthogonal to $ E+ \tilde E$ is compact or not. In the latter case one obtains the same result as in the non-compact case of eq. \eqref{annc}. In the case of compact $\tilde X^9$, the normalisation of the zero mode $\tilde p^9$ has been chosen such that its quantisation is the standard one
\begin{equation}
\tilde X^9  \rightarrow \tilde X^9 + 2 \pi R_{\perp} \qquad \implies \qquad \tilde p^9 = \frac{\tilde m_9}{ R_{\perp}}
\end{equation}
and the dependence on the electric field appears in $L_0$. Thus, in the case of a compact $\tilde X^9$, one obtains the following momentum sum contribution to the annulus amplitude 
\begin{align}
P_{\tilde m_9} =  \frac{1}{R_{\perp}} \sum_{\tilde m_9} e^{- \frac{\pi \tau_2\, \tilde m_9^2\, ||E+ \tilde E||^2}{2 R_{\perp}^2\, (||E+ \tilde E||^2- ||E \wedge \tilde E||^2)}}
\ .  \label{sum-oblique}
\end{align}
Notice that in the limit of parallel electric fields (thus setting $E\wedge \tilde E = 0$) the lattice sum reduces to the standard one,  as expected. Finally, the cylinder in the compact case, for oblique electric fields $E$ and $\tilde E$ can be written as
\begin{align}
\mathcal{A} =\frac{V_{10}}2 ||E+ \tilde E| |\int_0^\infty \frac{d\tau_2}{\tau_2^{9/2}}\sum_{\alpha,\beta}  c_{\alpha \beta} \frac{\vartheta \left[ \alpha \atop \beta\right] \left( i\epsilon \frac{i\tau_2}{2}\Big |  \frac{i \tau_2}{2}\right)}{\vartheta \left[\alpha \atop \beta\right] \left ( 0 \Big | \frac{i\tau_2}{2}\right)} \frac{\vartheta^4\left[\alpha \atop \beta \right] \left(0 \Big | \frac{i\tau_2}{2}\right)}{\eta^{12} \left(\frac{i\tau_2}{2}\right)}\frac{i \eta^3}{\vartheta_1\left ( i\epsilon \tfrac{i\tau_2}{2}|\tfrac{i\tau_2}{2}\right)}\,P_{\tilde m_9} \ .
\end{align}
Finally, let us consider the tree-level (transverse) channel properties of our cylinder amplitude. With the change of modular parameter in eq. \eqref{transverse} one finds
\begin{align}
\mathcal{\tilde A} =2^{-5}\frac{V_{10}}2 \sqrt{||E+ \tilde E| |^2-||E \wedge \tilde E||^2}\, \tilde R_9\int_0^\infty dl\, \sum_{\alpha,\beta}  c_{\alpha \beta} \frac{\vartheta \left[ \alpha \atop \beta\right] \left( i\epsilon  |  il \right)}{\vartheta \left[\alpha \atop \beta\right] \left ( 0  | il \right)} \frac{\vartheta^4\left[\alpha \atop \beta \right] \left(0 | i l\right)}{\eta^{12} \left(il\right)}\frac{i \eta^3(il)}{\vartheta_1\left ( i\epsilon |il\right)}\, \tilde P_{\tilde n_9}\ ,
\end{align}
where the dual lattice sum $\tilde P_{\tilde n_9}$ is obtained by Poisson summation
\begin{align}
\tilde P_{\tilde n_9} = \sum_{\tilde n_9} e^{-\frac{\pi l \, \tilde n_9^2 \tilde R_{\perp}^2\, (||E+ \tilde E||^2 - ||E \wedge \tilde E||^2) }{||E+ \tilde E||^2}} \ . 
\end{align}
As in the case of parallel electric fields, one needs to take into account the $\sinh \pi \epsilon$ factor from $\vartheta_1(i\epsilon|i l)$ in order to show consistency with the DBI interpretation. For oblique electric fields one can show the more general identity
\begin{equation}
\sinh \pi \epsilon = \frac{\sqrt{||E+ \tilde E||^2 - ||E \wedge \tilde E||^2}}{\sqrt{(1-||E||^2)(1-||\tilde E||^2)}} \ . 
\end{equation}
Thus one finds the following behaviour of the annulus amplitude in the transverse channel
\begin{align}
\mathcal{\tilde A} \sim \sqrt{(1-||E||^2)(1-||\tilde E||^2)} \ . 
\end{align}
This result is similar to the one obtained in the parallel case and it provides a non-trivial consistency check of the derived amplitudes in the presence of oblique electric fields.

\section{Energy loss of D-branes in electric fields}
\label{s8}
In what follows we denote the total energy loss by D-branes in an electric field by $W$, whereas the energy loss per (non-compact) spacetime volume will be
$w = W/V_D$, where $D$ is the number of non-compact dimensions. Schematically, our (cylinder) partition functions are of the form
\be
{\cal A} \equiv - i {\cal F} \equiv -i \int_0^{\infty} dt\, F(t)  \ , \label{loss1}
\ee
 where ${\cal F}$ is the vacuum energy, with $F(t)$ a {\it real} function.  In the presence of the electric field, the function $F$ has an infinity of poles 
 $t_k = \tau_{2,k} = 2 k/ |\epsilon|$ and the integral has an imaginary part calculated as the sum over all the residues, such that
 \be
 {\rm Im} \ {\cal F} =  {\rm Re} \ {\cal A} =  {\rm Im}  \int_0^{\infty} dt\,  F(t)  = \pi \sum_{k=1}^{\infty} {\rm Res} \ F(t_k) \ . \label{loss2}   
  \ee
 The probability of pair production is then given by the formula
 \be
 W = -2 \ {\rm Im} \ {\cal F} = - 2  \ {\rm Re} \ {\cal A}  \ . \label{loss3} 
  \ee
 It was  shown in \cite{Bachas:1992bh} that there is a general way to express the D-brane power loss. Slightly adapting it to our notation, it is given by
 \be
 w = \sum_{\rm states, S}  \frac{||E+ \tilde E||}{2 (2 \pi)^{D-1} \epsilon}  \sum_{k=1}^{\infty} (-1)^{(2 \alpha+1)( k+1)} \left(  \frac{|\epsilon|}{ k}\right)^{\frac{D}{2}} e^{- \frac{\pi k}{|\epsilon|} M_s^2} \ ,    \label{loss4} 
 \ee
 where $S$ denote all the states in the spectrum, including KK states and string oscillators, and $\alpha=0\ (1/2) $ for the NS (R) sector.
 
\begin{figure}[h]
\begin{center}
\includegraphics[width=\textwidth]{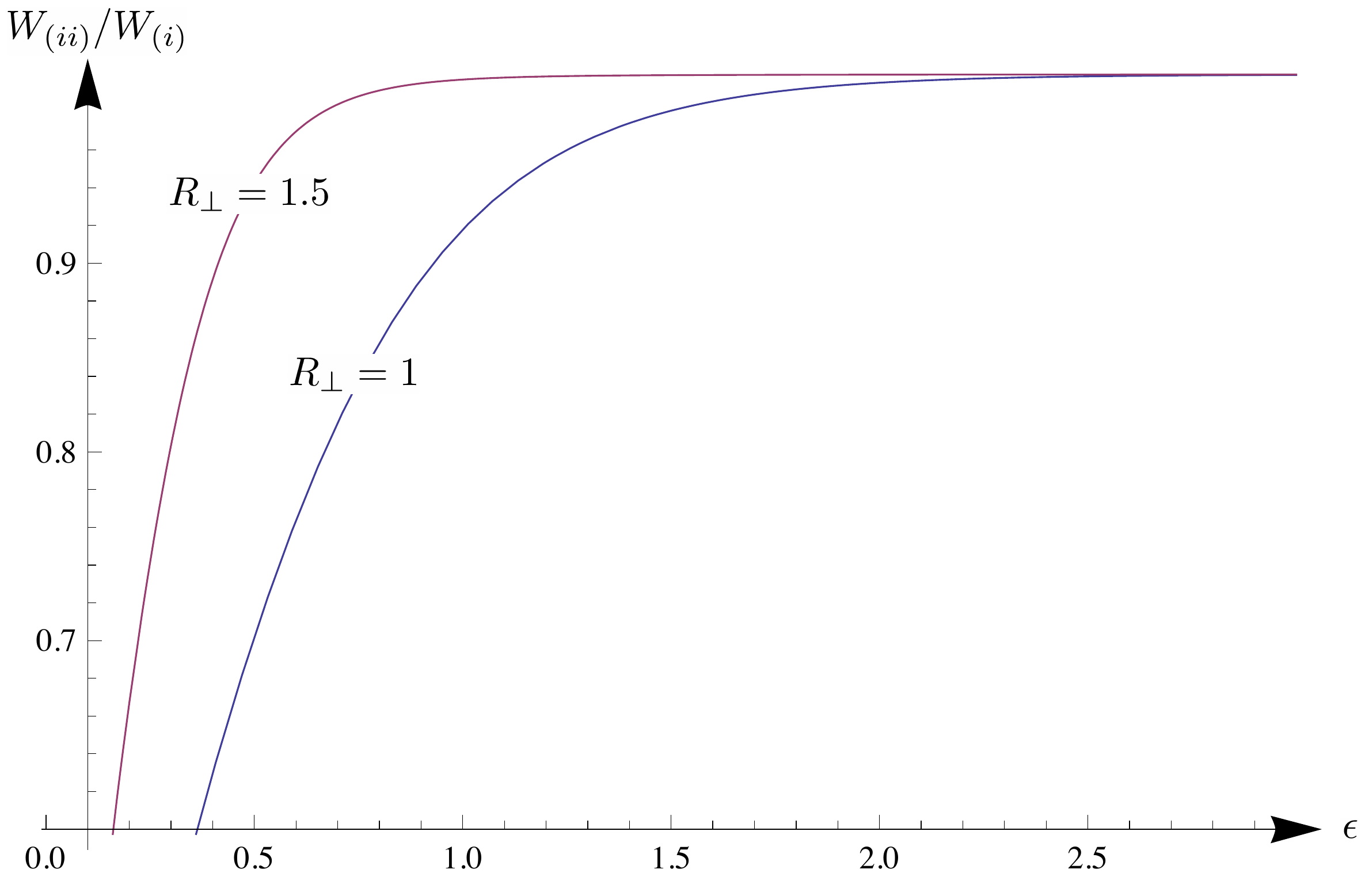}
\end{center}
\caption{We depict the dependence on $\epsilon$ of the ratio of the energy losses with $W_{(i)}$ given by eq. \eqref{loss6} with fixed $R_\perp = 1, 1.5$ and $W_{(ii)}: =  W_{(i)} \big  |_{R_\perp \rightarrow \infty}$ given in eq. \eqref{loss-nc}. The compact energy loss dominates over the non-compact one for any value of $\epsilon$. Asymptotically the ratio reaches the value 1 indicating the fact that both (compact and non-compact) cases diverge in the same way when the electric field reaches the limiting value.}\label{plot-epsilon}
\end{figure}

\begin{figure}[h]
\begin{center}
\includegraphics[width=\textwidth]{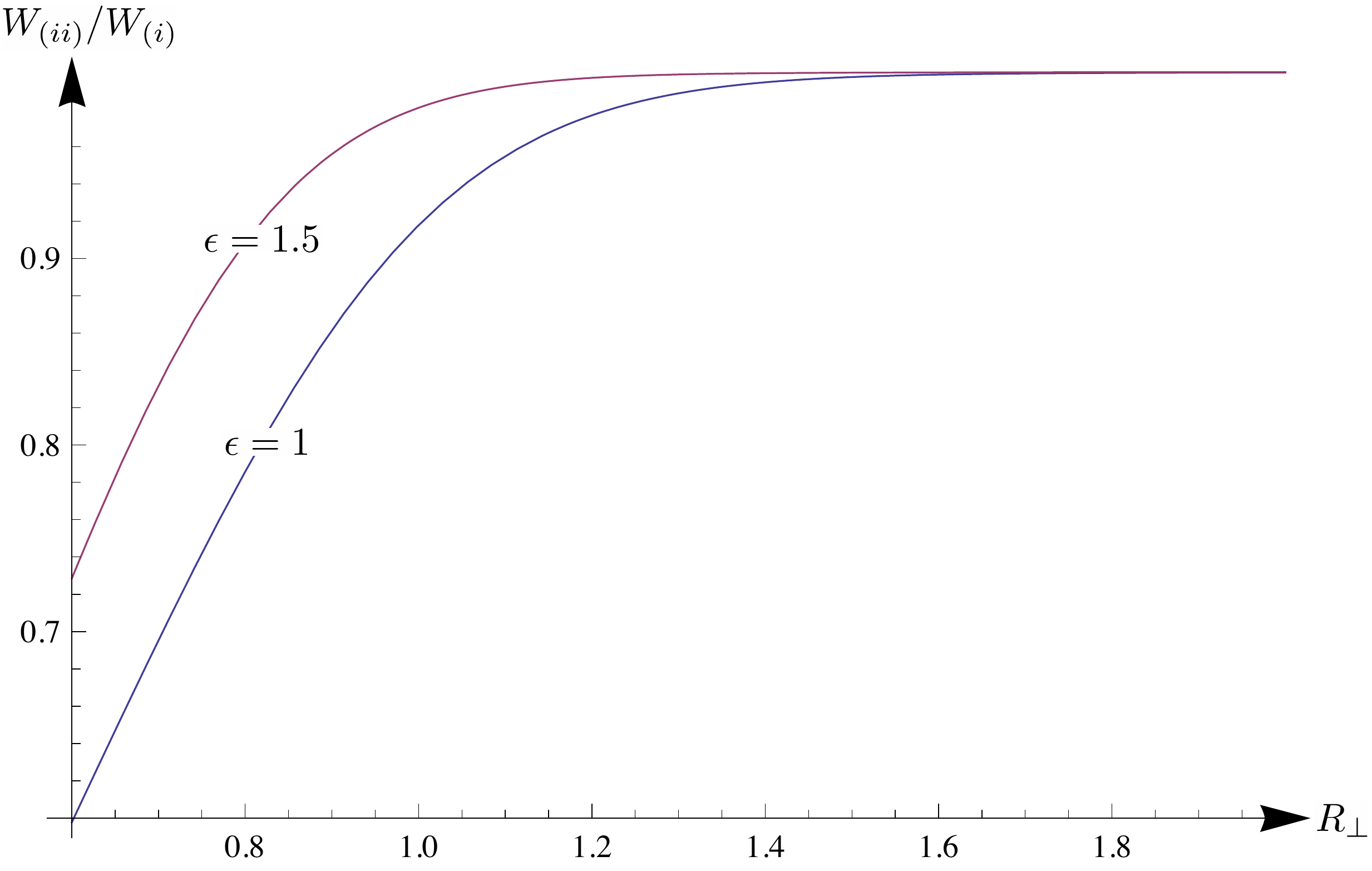}
\end{center}
\caption{We depict the dependence on $R_\perp$ of the ratio of the energy losses with $W_{(i)}$ given by eq. \eqref{loss6} with fixed $\epsilon =1, 1.5$ and $W_{(ii)}$ given by eq. \eqref{loss-nc} .The compact energy loss dominates over the non-compact one for any value of $R_\perp$. Asymptotically the ratio reaches the value 1 as expected from the definition of the non-compact energy loss $W_{(ii)}: =  W_{(i)} \big  |_{R_\perp \rightarrow \infty}$.}\label{plot-radius}
\end{figure}

 In the case of parallel electric fields, taking without loosing generality the case where the perpendicular coordinate to the electric field is compact, one can write the more
 convenient expression
  \be
 W =  \frac{ V_{10}\,  ||E+ \tilde E||}{(2 \pi)^{8} 2 R_{\perp} \epsilon}  \sum_{k=1}^{\infty}  \left(  \frac{|\epsilon|}{ k}\right)^{\frac{9}{2}} 
 \left(\frac{ (-1)^{k+1}V_8+  S_8}{\eta^8} \right) (\frac{i \tau_{2,k}}{2}= \frac{i k}{|\epsilon|})   \sum_{m_9 }e^{- \frac{\pi k}{|\epsilon|} \frac{m_9^2}{R_{\perp}^2}} \ ,    \label{loss5} 
 \ee
 where $V_{10}$ is spacetime volume and $V_8,S_8$ are $SO(8)$ characters (for their definition and properties, see e.g. \cite{orientifolds}).  Since $V_8=S_8$, only odd $k$ contribute to the energy loss,
 which can be therefore rewritten as 
 \be
 W =  \frac{ V_{10} ||E+ \tilde E||}{(2 \pi)^{8} R_{\perp} \epsilon}  \sum_{k=0}^{\infty}  \left(  \frac{|\epsilon|}{(2 k+1)}\right)^{\frac{9}{2}} 
 \left(\frac{\theta_2^4}{2 \eta^{12}} \right) (\frac{i (2k+1)}{|\epsilon|})   \sum_{m_9 }e^{-  \frac{\pi (2k+1)}{|\epsilon|} \frac{m_9^2}{R_{\perp}^2}} \ ,    \label{loss6} 
 \ee 
Notice that in the case of oblique sectors the expression of $W$ is of the same form with $\epsilon$ defined in eq. \eqref{e-oblique} and the (electric field dependent) lattice sum in eq. \eqref{sum-oblique}. 
In the following, we shall define $W_{(ii)}$ as the energy loss in the case of a non-compact orthogonal direction, {\it i.e.} we can write 
\begin{align}
W_{(ii)} = \lim_{R_\perp \rightarrow \infty} W_{(i)}\ ,
\end{align}
with the notation $W_{(i)}=W$ as the energy loss in the case of a compact orthogonal direction as given by eq. \eqref{loss6}. We now proceed to determine which of the $W_{(i)}$ and $W_{(ii)}$ is larger. For this purpose, we need the following inequality
\begin{align}
\frac{1}{R_\perp} \sum_{m\in \mathbb{Z}}e^{- \pi a \frac{m^2}{R_\perp^2}} & \geq \frac{1}{\sqrt{a}}\ , & \text{for all }\quad  a,R_\perp>0\ . \label{loss-ineq}
\end{align}
Furthermore,  the function on the left-hand side above decreases monotonically with the radius $R_\perp$ from $+\infty$ to $1/\sqrt{a}$ (one can see this by taking the derivative of the series after Poisson 
summation). Applying eq. \eqref{loss-ineq} term by term in the $k$-series of eq. \eqref{loss6}, with $a = (2k+1)/\epsilon$,  one can write
\begin{align}
W_{(i)} \geq \frac{V_{10}||E+ \tilde E||}{(2\pi)^8  \epsilon} \sum_{k=0}^\infty \left(\frac{|\epsilon|}{2k+1} \right)^{9/2} \frac{\vartheta_2^4}{\eta^{12}} \left( \tau=i \frac{2k+1}{|\epsilon|} \right) \left(\frac{|\epsilon|}{2k+1} \right)^{1/2} = W_{(ii)}\ , \label{loss-nc}
\end{align}
since every term in the series is positive. Thus we have shown that the energy loss in the compact case $W_{(i)}$ is always larger than the one in the non-compact case $W_{(ii)}$ for any finite positive values of $R_\perp$ and $\epsilon$. Since $W_{(i)/(ii)}$ are both positive quantities we have
\be
 0 \leq \frac{W_{(ii)}}{W_{(i)}} \leq 1 \ .
\ee
The value $0$ for the ratio is obtained in the limit $R_\perp \rightarrow 0$ or $\epsilon \rightarrow 0$ (it is not difficult to show that in this limit one has $W_{(ii)}/W_{(i)}\simeq |\epsilon|^{1/2} R_\perp$), whereas the value is $1$, in the limit $R_\perp \rightarrow \infty$ or $\epsilon \rightarrow \infty$. Indeed, we illustrate the dependence on $R_\perp$ and $\epsilon$ in Figures \ref{plot-epsilon}, \ref{plot-radius}.  
\\
\\
Hence, the energy loss by D-branes in the presence of a constant electric field is larger in the compact case such that it decreases monotonically with the radius of the orthogonal direction $R_\perp$ and reaches asymptotically the non-compact value. At the same time it increases with the electric field $\epsilon$, diverging when reaching the limiting value in such a way that the ratio non-compact/compact goes to 1. From the point of view of inflationary scenarios with D-branes we conclude that a larger (or infinite) radius would lead in principle to a larger number of e-folds.

\section{Conclusions}
\label{s9}
Our paper extends previously known results about open strings with background (constant) electric fields in two different ways: 

$\bullet$
First, and most important, we considered electric fields in compact spaces (tori) such that the direction of the electric field is at a generic angle with respect to (one of the) axes defining the torus lattice. The main result is that {\it the orientation of the electric field in the internal space has to be quantised}. Since this configuration is T-dual to D-branes moving with constant velocity in the internal space, we hope our results will be of some relevance for early cosmology and in particular inflation. We have given several derivations for the quantisation of the electric field direction stemming from the gauge invariance of Wilson lines, S-duality between electric and magnetic fields and the construction of quantum mechanical wave functions respecting the periodicities of the torus. The corresponding condition implies that the direction parallel to the electric field has to be compact. After T-duality, this implies that D-brane motion with constant velocity is periodic in the internal torus, with a periodicity $R_{\parallel} = \sqrt{p_2^2 R_2^2 + p_3^2 R_3^2}$, where $R_{2,3}$ are the internal radii of a rectangular torus and $p_{2,3}$ are integers, that can be parametrically large for large integers. This can have applications to inflation in string theory, particularly in string models with axion monodromy \cite{monodromy1}, where D-brane positions are natural inflaton candidates with large field excursions \cite{monodromy2}. However, the open string momenta allowed by the boundary conditions are always orthogonal to the electric field and may or may not be quantised depending on whether the orthogonal direction is compact or not. 

$\bullet$ 
The quantum mechanical analysis of the similar situation (by analytic continuation) involving a magnetic field yields the fact that the parallel  coordinate has to be compact as well, which can also be interpreted as the fact that the magnetic field is in the integral homology of the torus. For the case of the magnetic field and a different gauge, we could construct correct wavefunctions only if the perpendicular coordinate is also compact. If necessary, it is easy to see that this would further imply that the (absolute value squared of the) complex structure of the two torus where the magnetic field vector lies has to be fixed to a rational number. Such a condition, if necessary, can be of importance for the problem of moduli stabilisation.  However, we find such condition not to be necessary for the consistency of the cylinder string propagation (partition functions) and as such, we believe that it is an artefact of a special gauge choice.

$\bullet$ For the case of a particle/string in the presence of a magnetic field with corresponding vector pointing in an arbitrary direction in the yz-plane of a three torus, we showed that the degeneracy of the Landau levels is given by the greatest common divisor of the flux numbers in the xy- and xz-planes, a result that is important for model building in this framework.

$\bullet$
Second, we have performed the (covariant) quantisation of open strings with oblique electric fields in both non-compact and compact spaces, providing also the relevant amplitudes. The oblique sectors, which are always charged, appear naturally (only) in models involving several stacks of branes. They correspond to strings stretched between different branes with a non-zero angle between the background electric fields. As a result, the formulas for the algebra of string modes and for the electric field shift are somewhat more involved though preserving certain similarities with respect to the parallel case. In the non-relativistic limit (i.e. small electric fields) one recovers the results of parallel electric fields. This is obvious in the T-dual version where one has two branes moving with constant velocities in non-parallel directions. Going to the rest frame of one of the branes produces a Thomas precession effect which goes to zero in the non-relativistic limit. In view of this, the contribution of the oblique sectors can be important in studying the ultra-relativistic limit of such models. 

$\bullet$ 
Finally, we worked out the energy loss of D-branes in electric fields. It turns out that the result depends in a monotonically decreasing way on the length of the transverse coordinate to the electric field (which is by definition infinite if the corresponding direction is not periodic).  There is therefore a significant difference between the case of small length $R_{\perp}$ and the case of a large
(or infinite) one.  
 
\section*{Acknowledgements}
We thank Iosif Bena, Massimo Bianchi, Andrei Micu, Jihad Mourad and Augusto Sagnotti for enlightening discussions and comments. We are particularly grateful to Costas Bachas for collaboration in Sections 2,3,4 and numerous important discussions and comments. C.C. and G.P. are very grateful to CPHT - Ecole Polytechnique and C.C. is also grateful to the INFN Section and the Physics Department of the University of Rome ``Tor Vergata'' for the kind hospitality during various stages of this work. C.C. was supported by the grant PN 16 42 01 01/2016. E.D. was supported in part by the ``Agence Nationale de la Recherche" (ANR) grant Black-dS-String.  G.P. was supported in part by the ``String Theory and Inflation'' Uncovering Excellence Grant of the University of Rome ``Tor Vergata", CUP E82L15000300005, and by the MIUR PRIN Contract 2015MP2CX4 ``Non-perturbative Aspects Of Gauge Theories And String''.

\appendix
\section{Fermions}
Worldsheet supersymmetry implies that the fermionic coordinates $\Psi_{L,R}^\mu$ satisfy the same boundary conditions as the derivatives of the bosonic ones $\partial X_L^\mu$ and $\partial X_R^\mu$ up to sign depending on the sector (NS or R)
\begin{align}
\Psi_L^\mu(\tilde \tau+2 \pi) &= (-1)^k (\Lambda_2 \Lambda_1)^\mu{}_\nu\, \Psi_L^\nu(\tilde \tau) \ ,\\
\Psi_R^\mu(\tilde \tau) & = \Lambda_1^\mu{}_\nu \, \Psi_L^\nu (\tilde \tau)\ ,
\end{align}
where $k=0,1$ for periodic or anti-periodic boundary conditions. The mode expansions can then be easily written for the coordinates $\Psi^a:= (C^{-1})^a{}_\mu \, \Psi^\mu$ as follows
\begin{align}
\Psi_L^a(\sigma_+) &= \sum_{n\in \mathbb{Z}+k/2} b_{n+i\epsilon_a}^a\, e^{-i(n+i\epsilon_a)\sigma_+}\ , && \Psi_R^a(\sigma_-) &= \sum_{n\in \mathbb{Z}+k/2} \Lambda_1^a{}_b\, b^b_{n+i \epsilon_b}\, e^{-i(n+i \epsilon_b)\sigma_-}
\end{align}
which after canonical quantisation leads to the usual algebra for the oscillator modes
\begin{align}
\left[b^a_{n+i\epsilon_a}, b_{m+i\epsilon_b}^b\right] = i\,\eta^{ab}\, \delta_{m+n,0}\, \delta_{\epsilon_a + \epsilon_b,0}\ .
\end{align}
It then follows immediately that the contribution of the fermions to the annulus amplitude (in both the parallel and oblique case) is of the form
\begin{align}
\mathcal{A}_{f} \sim \sum_{\alpha,\beta}  c_{\alpha \beta} \frac{\vartheta \left[ \alpha \atop \beta\right] \left( i\epsilon \frac{i\tau_2}{2}\Big |  \frac{i \tau_2}{2}\right)}{\eta \left(\frac{i\tau_2}{2}\right)} \frac{\vartheta^4\left[\alpha \atop \beta \right] \left(0 \Big | \frac{i\tau_2}{2}\right)}{\eta^{4} \left(\frac{i\tau_2}{2}\right)}
\end{align}
where $\epsilon$ is given by eq. \eqref{epspar} in the parallel case and by eq. \eqref{e-oblique} in the oblique case. The coefficients $c_{\alpha\beta} :=(-1)^{2\alpha + 2 \beta + 4 \alpha \beta} $, with $\alpha,\beta = 0,1/2$, take into account the usual summation over the spin structures.

\end{document}